\documentclass[10pt,english,floatfix,nofootinbib,superscriptaddress,aps,prd,preprint, twocolumn]{revtex4}
\usepackage[utf8]{inputenc}
\usepackage{float}
\usepackage{array}
\usepackage{lipsum}
\usepackage{bbm}
\usepackage{dsfont}
\usepackage{graphicx}
\usepackage{amsmath}
\usepackage{tikz}
\usepackage{multirow}
\usepackage{braket}
\usepackage{bbm}
\usetikzlibrary{quotes,angles}
\usetikzlibrary{arrows} 
\usetikzlibrary{decorations.markings}
\usepackage{graphicx}
\usepackage[english]{babel}
\usepackage{color}
\usepackage{subfig}
\usepackage{caption}
\usepackage{tensor}
\usepackage{esint}
\usepackage[dvips]{epsfig}
\usepackage[dvips]{graphicx}
\usepackage{float}
\usepackage{units}
\usepackage{textcomp}
\usepackage{mathrsfs}
\usepackage{amsmath}
\usepackage[makeroom]{cancel}
\usepackage{amssymb}
\usepackage{amsbsy}
\usepackage{amsfonts}
\usepackage{amssymb,mathrsfs,xcolor}
\usepackage{esint}
\usepackage{array}
\usepackage{graphicx}
\usepackage{slashed}

\newcommand{\ie}{\begin{equation}}
\newcommand{\fe}{\end{equation}}
\newcommand{\se}{\begin{eqnarray}}
\newcommand{\ff}{\end{eqnarray}}

\usepackage{hyperref}
\hypersetup{colorlinks,breaklinks,
			citecolor=[rgb]{0,0.0,1.0},
            urlcolor=[rgb]{0.0,0.0,1.0},
            linkcolor=[rgb]{0,0.5,0.9}}

\begin{document}

\title{Neutrino oscillations in a Kalb–Ramond black hole background}

\author{Yuxuan Shi}
\email{shiyx2280771974@gmail.com}
\affiliation{Department of Physics, East China University of Science and Technology, Shanghai 200237, China}

%%%%%%%%%%%%%%%%%%%%%%%%%%%%%%%%%%%%%%%%%%%%%%%%%%%%%%%%%%%%%%%%%%%%%%%%%%%%%%%%%%%%%%%%%%%%%%%%%%%%%%%%

\author{A. A. Ara\'{u}jo Filho}
\email{dilto@fisica.ufc.br}
\affiliation{Departamento de Física, Universidade Federal da Paraíba, Caixa Postal 5008, 58051--970, João Pessoa, Paraíba,  Brazil.}

%%%%%%%%%%%%%%%%%%%%%%%%%%%%%%%%%%%%%%%%%%%%%%%%%%%%%%%%%%%%%%%%%%%%%%%%%%%%%%%%%%%%%%%%%%%%%%%%%%%%%%%%

\author{K. E. L. de Farias }
\email{klecio.lima@uaf.ufcg.edu.br}

\affiliation{Departamento de Física, Universidade Federal de Campina Grande Caixa Postal 10071, 58429-900 Campina Grande, Paraíba, Brazil.}

%%%%%%%%%%%%%%%%%%%%%%%%%%%%%%%%%%%%%%%%%%%%%%%%%%%%%%%%%%%%%%%%%%%%%%%%%%%%%%%%%%%%%%%%%%%%%%%%%%%%%%%%

\author{V. B. Bezerra}
\email{valdir@fisica.ufpb.br}
\affiliation{Departamento de Física, Universidade Federal da Paraíba, Caixa Postal 5008, 58051--970, João Pessoa, Paraíba,  Brazil.}

%%%%%%%%%%%%%%%%%%%%%%%%%%%%%%%%%%%%%%%%%%%%%%%%%%%%%%%%%%%%%%%%%%%%%%%%%%%%%%%%%%%%%%%%%%%%%%%%%%%%%%%%

\author{Amilcar R. Queiroz}
\email{amilcarq@df.ufcg.edu.br}

\affiliation{Departamento de Física, Universidade Federal de Campina Grande Caixa Postal 10071, 58429-900 Campina Grande, Paraíba, Brazil.}

%%%%%%%%%%%%%%%%%%%%%%%%%%%%%%%%%%%%%%%%%%%%%%%%%%%%%%%%%%%%%%%%%%%%%%%%%%%%%%%%%%%%%%%%%%%%%%%%%%%%%%%%%%%%%%%%%%%%%%%%%%%%%%%%%%%%%%%%%%%%%%%%%%%%%%%%%%%%%%%%%%%%%%%%%%%%%%%%%%%%%%%%%%%%%%%%%%%%%%%%%%%%%%%%%%%%%%%%%%%%%%%%%%%%%%%%%%%%%%%%%%%%%%%%%%%%%%%%%%%%%%%%%%%%%%%%%%%%%%%%%%%%%%%%%%%%%%%%%%%%%%%%%%%%%%%%%%%%%%%%%%%%%%%%%%%%%%%%%%%%%%%%%%%%%%%%%%%%%%%%%%%%%%%%%%%%%%%%%%%%%%%%%%%%%%%%%%%

\date{\today}

\begin{abstract}

The analysis examines how neutrinos behave when their trajectories unfold around a black hole sourced by a Kalb--Ramond field, where spontaneous Lorentz symmetry breaking reshapes the surrounding geometry. Instead of following the conventional order, the study focuses first on the observable consequences: alterations in the neutrino--antineutrino annihilation energy output, shifts in the oscillation phase accumulated along the path, and distortions in flavor conversion probabilities induced by gravitational lensing. These features are then tied to the Lorentz--violating spacetime structure, which governs the propagation of the neutrinos. Numerical simulations are carried out for both two-- and three--flavor descriptions, with normal and inverted mass orderings.

\end{abstract}

\maketitle

\tableofcontents

%%%%%%%%%%%%%%%%%%%%%%%%%%%%%%%%%%%%%%%%%%%%%%%%%%%%%%%%%%%%%%%%%%%%%%%%%%%%%%%%%%%%%%%%%%%%%%%%%%%%%%%%%%%%%%%%%%%%%%%%%%%%%%%%%%%%%%%%%%%%%%%%%%%%%%%%%%%%%%%%%%%%%%%%%%%%%%%%%%%%%%%%%%%%%%%%%%%%%%%%%%%%%%%%%%%%%%%%%%%%%%%%%%%%%%%%%%%%%%%%%%%%%%%%%%%%%%%%%%%%%%%%%%%%%%%%%%%%%%%%%%%%%%%%%%%%%%%%%%%%%%%%%%%%%%%%%%%%%%%%%%%%%%%%%%%%%%%%%%%%%%%%%%%%%%%%%%%%%%%%%%%%%%%%%%%%%%%%%%%%%%%%%%%%%%%%%%%%%%%%%%%%%%%%%%%%%%%%%%%%%%%%%%%%%%%%%%%%%%%%%%%%%%%%%%%%%%%%%%%%%%%%%%%%%%%%%%%%%%%%%%%%%%

\section{Introduction}

The invariance of physical laws under transformations between inertial observers is a central feature of Lorentz symmetry. Despite the extensive experimental support for this principle, several frameworks devised for regimes near or beyond the Planck scale allow for departures from exact Lorentz invariance. A wide range of studies has explored these possibilities \cite{int6,int5,int4,int2,heidari2023gravitational,int1,araujo2023thermodynamics,int3}, motivated by the prospect that new physics may surface through tiny deviations from standard relativistic behavior.
Proposals that accommodate Lorentz violation usually fall into two broad categories. In one class, the symmetry is broken at the level of the field equations: the Lagrangian incorporates terms that already select preferred directions or frames, and observable quantities inherit this asymmetry. A different mechanism operates when the underlying equations preserve Lorentz covariance while the vacuum configuration does not. In this scenario, a nontrivial background field develops a vacuum expectation value that singles out specific directions, triggering spontaneous symmetry breaking with noteworthy physical implications \cite{Bluhm:2023kph,bluhm2006overview,bluhm2005spontaneous,Bluhm:2019ato,bluhm2008spontaneous,bluhm2008constraints}.

Investigations of scenarios where the vacuum configuration departs from perfect Lorentz symmetry frequently rely on the framework provided by the Standard Model Extension (SME) \cite{colladay1998lorentz,11,12,10,13,9,Amarilo:2023wpn}. Within this setting, one prominent avenue involves vector fields that acquire a fixed background value. These constructions, broadly referred to as bumblebee models, posit a vector sector whose vacuum expectation value selects preferred directions in spacetime, thereby modifying the interaction patterns of matter and fields. Rather than treating Lorentz violation as an explicit deformation of the fundamental equations, this approach attributes the symmetry departure to the vacuum itself, which reshapes the local structure of spacetime. This mechanism has been examined in a variety of gravitational environments, where it influences thermal properties, particle production processes, and other phenomena associated with black holes and curved backgrounds \cite{anacleto2018lorentz,aa2021lorentz,araujo2022thermal,paperrainbow,aa2022particles,araujo2021higher,Liu:2022dcn}.

Moreover, Early attempts to incorporate spontaneous Lorentz violation into gravitational backgrounds yielded the first static, spherically symmetric configuration in bumblebee gravity, as presented in Ref.~\cite{14}. This initial geometries later served as starting points for a variety of generalizations. Schwarzschild--type metrics were reformulated to include the influence of the bumblebee field, and subsequent work introduced their counterparts in (A)dS settings, producing modified Schwarzschild–(A)dS spacetimes \cite{20}. Other developments proceeded in different directions. A non--commutative deformation of gauge gravity was implemented for the bumblebee scenario, giving rise to a non--commutative version of the black hole solution \cite{AraujoFilho:2025rvn}. Parallel to these efforts, the theory was revisited within the \textit{metric--affine} formalism: first through a static configuration \cite{Filho:2022yrk}, and later through an axisymmetric extension \cite{AraujoFilho:2024ykw}.

Another line of research addressing departures from exact Lorentz symmetry relies on an antisymmetric tensor sector rather than on a vector field. The Kalb--Ramond field, a rank--two tensor naturally emerging in bosonic string theory \cite{43}, can acquire a vacuum configuration that selects preferred directions once it interacts nonminimally with the gravitational sector \cite{maluf2019antisymmetric,42}. In this setting, the vacuum expectation value of the antisymmetric field triggers spontaneous Lorentz violation.

Black hole geometries supported by this mechanism have been constructed and analyzed from several perspectives. A static, spherically symmetric solution furnished the first concrete example \cite{yang2023static}, and subsequent work explored the behavior of particles and fields in its vicinity \cite{45}. Over time, many aspects of these backgrounds have been examined: quasinormal oscillations \cite{araujo2024exploring}, greybody transmission bounds and spectra \cite{guo2024quasinormal}, lensing and strong--field optical features \cite{junior2024gravitational}, limits on symmetry-breaking scales \cite{junior2024spontaneous}, circular trajectories and QPOs \cite{jumaniyozov2024circular}, accretion dynamics described through Vlasov gas models \cite{jiang2024accretion} and particle creation \cite{AraujoFilho:2024ctw,araujo2025does}.

Charged configurations have also been constructed \cite{duan2024electrically}, inspiring additional studies examining thermodynamic behavior, radiative processes, and geometric effects \cite{al-Badawi:2024pdx,chen2024thermal,Zahid:2024ohn,aa2024antisymmetric,heidari2024impact}. A separate development introduced a non-commutative deformation of the Kalb--Ramond black hole \cite{AraujoFilho:2025jcu}. Slowly rotating versions of this geometry have likewise attracted attention, with analyses of shadow contours \cite{Liu:2024lve} and quasinormal spectra for scalar, vector, and tensor perturbations \cite{Deng:2025atg}. Furthermore, configurations endowed with global--monopole parameters have been constructed in this framework \cite{Belchior:2025xam}.

A subsequent development appeared in Ref.~\cite{Liu:2024oas}, where a different static black hole geometry was introduced
\begin{align}
\label{maetriiccc}
\mathrm{d}s^{2} = & -\left(1-\dfrac{2M}{r}\right)\mathrm{d}t^{2} + (1-\ell)\left(1-\dfrac{2M}{r}\right)^{-1}\mathrm{d}r^{2}\notag\\
& + r^{2}\mathrm{d}\theta^{2} + r^{2} \sin^{2}\mathrm{d}\varphi^{2}.
\end{align} 
In this background, subsequent analyses examined how the geometry affects entanglement loss \cite{liu2024lorentz} as well as the mechanism of particle production \cite{AraujoFilho:2024ctw}.

Neutrinos occupy a singular position among the fermions of the Standard Model. While quark mixing and meson oscillations have long been established phenomena, only neutrinos can maintain quantum coherence over scales ranging from laboratory baselines to interstellar distances, a consequence of their extremely small masses and the misalignment between flavor and mass sectors \cite{Pontecorvo1,maki1962remarks,Pontecorvo2}. This peculiar behavior has made them exceptionally responsive to potential imprints of physics beyond the Standard Model and to gravitational influences encountered along their trajectories \cite{neu42,neu44,neu43}. A remarkable aspect behind this sensitivity lies in the fact that neutrinos created in weak processes belong to flavor eigenstates, whereas their propagation is governed by mass eigenstates. Since these two bases are not identical, a neutrino emitted with a definite flavor evolves as a coherent superposition of mass modes. The interference among these components reshapes the flavor content as the particle travels, producing the characteristic oscillatory modulation observed in experiments and astrophysical environments alike \cite{neu41,neu40,neu39}.

When neutrino propagation is examined in flat spacetime, the oscillatory behavior does not hinge on the absolute values of the masses but instead on how the squared masses differ from one another. The relevant parameters are introduced through
$\Delta m^2_{ij} = m_i^{2} - m_j^{2}$, which quantify the separations between the mass eigenstates. Experimental analyses typically focus on the magnitudes $|\Delta m^2_{21}|$, $|\Delta m^2_{31}|$, and $|\Delta m^2_{23}|$. Because the oscillation probabilities are governed solely by these differences, the phenomenon offers no direct access to the overall neutrino mass scale; only the spacing between the mass levels can be inferred \cite{neu45}.

When neutrinos traverse curved backgrounds, their oscillatory behavior acquires features that do not arise in flat spacetime. The accumulated phase no longer depends exclusively on the gaps between the squared masses; rather, gravitational curvature introduces additional contributions that may carry sensitivity to the absolute mass scale \cite{Alloqulov:2024sns,Chakrabarty:2023kld,Shi:2024flw,Shi:2025rfq,AraujoFilho:2025rzh}. These modifications reshape the standard phase relations and become increasingly relevant for neutrinos originating from energetic astrophysical or cosmological environments \cite{Shi:2023kid}.

As a neutrino beam moves through regions where the geometry differs from Minkowski space, the curvature--induced phase shift intertwines information about the particle’s intrinsic parameters with details of the spacetime it crosses. Deviations between the observed flavor composition and the expectations from the usual flat--spacetime oscillation formulae can therefore act as a signature of gravitational influence. This perspective allows one to use flavor evolution simultaneously as a probe of the gravitational setting and as a means to investigate the structure of the neutrino mass spectrum \cite{neu48,neu52,neu47,neu50,neu46,neu51,Shi:2023hbw,neu49,neu53}.

The evolution of neutrino flavors is intimately connected to the spacetime through which they propagate. The oscillation phase accumulated along a neutrino worldline carries the imprint of the surrounding geometry \cite{neu54}, a consequence that naturally emerges once the process is formulated in geometric terms. In regions where curvature is pronounced—particularly in the vicinity of compact objects—the gravitational field can deflect neutrino trajectories and even cause nearby paths to converge. Such lensing effects modify the relative propagation of the mass eigenstates and, consequently, reshape the interference pattern responsible for oscillations. As a result, the predicted transition probabilities acquire corrections linked to these gravitational distortions \cite{neu53,Shi:2025rfq,Shi:2024flw,Shi:2025plr}.

A growing body of work has examined how the maintenance of quantum coherence in neutrino oscillations is influenced by curved backgrounds, with particular attention given to distortions produced by gravitational lensing \cite{neu57,neu56,neu58}. When the spacetime originates from a rotating object, the situation becomes even more intricate. The angular momentum of the gravitational source modifies the phase accumulated by the propagating neutrinos, an effect highlighted in Swami’s analysis. Depending on the rotational geometry, these corrections can increase or diminish the expected flavor--transition probabilities. Such phenomena are not limited to extreme astrophysical settings; even gravitational fields generated by stellar--mass objects can imprint noticeable rotational contributions to the evolution of neutrino flavors \cite{neu59}.

Flavor evolution has also been examined in settings where the background geometry departs from spherical symmetry. Axially deformed spacetimes, characterized by a deformation parameter $\gamma$, provide a notable example. Here, the metric of a static and asymptotically flat configuration is reshaped by the value of $\gamma$, and this geometric modification feeds directly into the neutrino phase accumulated along the trajectory. Because of this altered phase structure, neutrino oscillations in such backgrounds can display a sensitivity to the absolute mass scale—an effect absent in standard flat--spacetime analyses. In other words, the emergence of this mass dependence shows how deviations from spherical symmetry can substantially influence flavor--transition behavior \cite{neu60}.

In this manner, this work investigates neutrino propagation in the exterior region of a black hole supported by a Kalb--Ramond background, where spontaneous Lorentz violation modifies the underlying spacetime. The discussion is organized around the physical manifestations of this deformation rather than beginning with the geometry itself. Observable quantities take center stage: variations in the energy produced through $\nu\bar\nu$ annihilation, changes in the oscillation phase accumulated along neutrino paths, and lensing--induced modifications to flavor--transition probabilities.
Afterwards, these effects linked to the structure of the Lorentz--violating metric responsible for guiding the trajectories. The analysis incorporates numerical evaluations of flavor evolution in both two-- and three--flavor frameworks, considering normal and inverted neutrino mass hierarchies.

%%%%%%%%%%%%%%%%%%%%%%%%%%%%%%%%%%%%%%%%%%%%%%%%%%%%%%%%%%%%%%%%%%%%%%%%%%%%%%%%%%%%%%%%%%%%%%%%%%%%%%%%%%%%%%%%%%%%%%%%%%%%%%%%%%%%%%%%%%%%%%%%%%%%%%%%%%%%%%%%%%%%%%%%%%%%%%%%%%%%%%%%%%%%%%%%%%%%%%%%%%%%%%%%%%%%%%%%%%%%%%%%%%%%%%%%%%%%%%%%%%%%%%%%%%%%%%%%%%%%%%%%%%%%%%%%%%%%%%%%%%%%%%%%%%%%%%%%%%%%%%%%%%%%%%%%%%%%%%%%%%%%%%%%%%%%%%%%%%%%%%%%%%%%%%%%%%%%%%%%%%%%%%%%%%%%%%%%%%%%%%%%%%%%%%%%%%%%%%%%%%%%%%%%%%%%%%%%%%%%%%%%%%%%%%%%%%%%%%%%%%%%%%%%%%%%%%%%%%%%%%%%%%%%%%%%%%%%%%%%%%%%%%%%%%%%%%%%%%%%%%%%

\section{Energy release produced by particle-antiparticle annihilation }

In this part of the analysis, the quantity of interest was the energy delivered to the surrounding region by neutrino--antineutrino interactions in the gravitational geometry defined by the Lorentz--violating parameter $\ell$ of Eq.~(\ref{maetriiccc}). Rather than beginning with the background itself, the discussion was organized around the physical channel responsible for transferring energy: the annihilation of the two species. Once this mechanism was established as the dominant contributor, the corresponding deposition rate—evaluated per unit volume and per unit time—was introduced through the standard expression \cite{Salmonson:1999es}:
\ie
\dfrac{\mathrm{d}\mathrm{E}(r)}{\mathrm{d}t\,\mathrm{d}V}=2 \, \mathrm{K} \,\mathrm{G}_{f}^{2}\, \mathrm{f}(r)\iint
n(\varrho_{\nu})n(\varrho_{\overline{\nu}})
(\varrho_{\nu} + \varrho_{\overline{\nu}})
\varrho_{\nu}^{3}\varrho_{\overline{\nu}}^{3}
\mathrm{d}\varrho_{\nu}\mathrm{d} \varrho_{\overline{\nu}}
\fe
in which
\ie
\mathrm{K} = \dfrac{1}{6\pi}(1\pm4\sin^{2}\vartheta_{W}+8\sin^{4} \vartheta_{W}).
\fe

Using the conventional choice $\sin^{2}\vartheta_{W} = 0.23$ for the Weinberg angle, the weak interaction sector fixes the coefficients that govern each neutrino--antineutrino annihilation channel. With these parameters specified, the expressions reported in \cite{Salmonson:1999es} furnish the corresponding energy deposition rates, showing how the flavor involved and the strength of the weak coupling determine the amount of energy delivered through annihilation
\ie
\mathrm{K}(\nu_{\mu},\overline{\nu}_{\mu}) = \mathrm{K}(\nu_{\tau},\overline{\nu}_{\tau})
=\dfrac{1}{6\pi}\left(1-4\sin^{2}\vartheta_{W} + 8\sin^{4}\vartheta_{W}\right),
\fe
and also
\ie
\mathrm{K}(\nu_{e},\overline{\nu}_{e})
=\dfrac{1}{6\pi}\left(1+4\sin^{2}\vartheta_{W} + 8\sin^{4}\vartheta_{W}\right).
\fe

The starting point for assigning an energy–release contribution to each neutrino--antineutrino pair is the specification of the weak--interaction parameters. Once the Fermi constant is fixed at $\mathrm{G}_f = 5.29 \times 10^{-44}\,\text{cm}^2\,\text{MeV}^{-2}$ and the Weinberg angle is taken through $\sin^{2}\vartheta_W = 0.23$, the coefficients that differentiate the various flavor channels follow immediately. With these constants in place, one can obtain the individual deposition rates. Carrying out the angular integration then leads to the compact formula reported in \cite{Salmonson:1999es}
\begin{align}
\mathrm{f}(r)&=\iint\left(1-\bm{\overset{\nsim}{\Omega}_{\nu}}\cdot\bm{\overset{\nsim}{\Omega}_{\overline{\nu}}}\right)^{2}
\mathrm{d}    \overset{\nsim}{\Omega}_{\nu}\mathrm{d}\overset{\nsim}{\Omega}_{\overline{\nu}}\notag\\
&=\dfrac{2\pi^{2}}{3}(1 - x)^{4}\left(x^{2} + 4x + 5\right)
\end{align}
where we have
\ie
x = \sin\vartheta_{r}.
\fe

For a fixed radial position $r$, the geometry sets the ``inclination'' of the particle trajectories: the quantity $\vartheta_{r}$ characterizes the departure from the local tangential direction defined by the circular orbit at that radius. Instead of beginning with this angle, the discussion may be framed by first specifying the orientation of the fluxes. The propagation of neutrinos and antineutrinos is encoded in the unit vectors $\overset{\nsim}{\Omega}_{\nu}$ and $\overset{\nsim}{\Omega}_{\overline{\nu}}$, with their respective solid--angle elements $\mathrm{d}\overset{\nsim}{\Omega}_{\nu}$ and $\mathrm{d}\overset{\nsim}{\Omega}_{\overline{\nu}}$ entering the angular integrations that define the annihilation contribution. When the medium is assumed to have thermalized at temperature $\Tilde{T}$, both species occupy their phase space according to Fermi--Dirac distributions, written as $n(\varrho_{\nu})$ and $n(\varrho_{\overline{\nu}})$ in the notation of \cite{Salmonson:1999es}
\ie
n(\varrho_{\nu}) = \frac{2}{h^{3}}\dfrac{1}{e^{\left({\frac{\varrho_{\nu}}{k \, \Tilde{T}}}\right)} + 1}.
\fe

The description of the energy delivered by neutrino--antineutrino annihilation may be organized around the fundamental constants that govern the quantum–thermal character of the process. Once Planck’s constant $h$ and Boltzmann’s constant $k$ are fixed, these quantities set the scale for the expression that measures the deposited energy per unit volume and per unit time. The resulting rate, obtained after incorporating the statistical and quantum ingredients of the system, follows the formulation presented in \cite{Salmonson:1999es}
\ie
\frac{\mathrm{d}\mathrm{E}}{\mathrm{d}t\mathrm{d}V} = \frac{21\zeta(5)\pi^{4}}{h^{6}}\mathrm{K} \, \mathrm{G}_{f}^{2} \, \mathrm{f}(r)(k \, \Tilde{T})^{9}.
\fe
The starting point for discussing energy exchange in the vicinity of compact objects is the radial dependence of the quantity $\mathrm{d}E/(\mathrm{d}t\,\mathrm{d}V)$, which serves as the measure of how much energy is deposited within a given spacetime region \cite{Salmonson:1999es}. Instead of framing it as a transformation rate, the emphasis may be placed on the parameters that dictate its variation with radius. Among these, the temperature profile $\Tilde{T}(r)$ is essential, since it determines the thermal state of the medium through the entire domain under consideration \cite{Salmonson:1999es}.

The local temperature registered by an observer at radius $r$ does not coincide with the emission temperature due to gravitational redshift. This effect is encoded in the condition
$\Tilde{T}(r)\,\sqrt{-\mathrm{g}_{tt}(r)} = \text{constant}$, which signals how the geometry modifies thermal measurements \cite{Salmonson:1999es}. Evaluating this condition at the neutrinosphere, located at $r = \mathrm{R}$, fixes the temperature that characterizes the neutrino outflow \cite{Salmonson:1999es}
\ie
\Tilde{T}(r)\sqrt{-\mathrm{g}_{tt}(r)} = \Tilde{T}(\mathrm{R})\sqrt{-\mathrm{g}_{tt}(\mathrm{R})}.
\fe  
The quantity $\mathrm{R}$ identifies the radius of the emitting object that generates the gravitational field in which the neutrinos propagate. Rather than beginning with this geometric scale, the calculation is streamlined by first recasting the temperature profile in terms of the redshift relation established previously. Once this condition is imposed, the temperature measured at any radius can be expressed in a form suitable for substitution into the luminosity integral. Incorporating the gravitational redshift in this way leads directly to the expression for the neutrino luminosity reported in \cite{Salmonson:1999es}
\ie
\Tilde{\mathcal{L}}_{\infty} = -\mathrm{g}_{tt}(\mathrm{R})L(\mathrm{R}).
\fe

The power carried away by each neutrino species, when assessed at the neutrinosphere, follows from the standard expression used at that radius. Once the emission temperature at $r=\mathrm{R}$ is fixed, the luminosity for an individual flavor is obtained through the relation provided in \cite{Salmonson:1999es}
\ie
\Tilde{\mathcal{L}}(\mathrm{R}) = 4 \pi \mathrm{R}_{0}^{2}\dfrac{7}{4}\dfrac{a\,c}{4}\Tilde{T}^{4}(\mathrm{R}).
\fe

The constants $c$ and $a$ enter as fixed coefficients in the luminosity expression. After specifying them, the next step is to express the temperature at radius $r$ in its redshifted form, since measurements made away from the neutrino-sphere must reflect the local geometry. Implementing this requirement leads to the relation employed in \cite{Salmonson:1999es}:

\begin{align}
\frac{\mathrm{d}\mathrm{E}(r)}{\mathrm{d}t \, \mathrm{d}V} & = \dfrac{21\zeta(5)\pi^{4}}{h^{6}}
\mathrm{K} \, \mathrm{G}_{f}^{2} \, k^{9}\left(\dfrac{7}{4}\pi a\,c\right)^{-\frac{9}{4}}\\
& \times \Tilde{\mathcal{L}}_{\infty}^{\frac{9}{4}}\mathrm{f}(r)
\left[\dfrac{\sqrt{-\mathrm{g}_{tt}(\mathrm{R})}}{-\mathrm{g}_{tt}(r)}\right]^{\frac{9}{2}} \mathrm{R}^{-\frac{9}{2}}.
\end{align}

The symbol $\zeta(s)$ appearing in the preceding formula designates the Riemann zeta function. For real arguments exceeding unity, its standard introduction is through an infinite series that serves as its defining representation:
\begin{align}
\zeta(s) = \sum_{n=1}^{\infty} \frac{1}{n^s}. 
\end{align}

The quantity describing how much energy is deposited at a given point is governed both by the radial position and by the geometric data set by the compact object’s metric. Rather than starting from this observation, the evaluation of the total radiated energy may be organized by first accumulating the local deposition rate over time in the curved background. The angular sector enters through the function $\mathrm{f}(r)$, whose construction hinges on the variable $x$ introduced earlier for the angular integration. Its treatment, refined for the present context, leads to the sequence of steps outlined in \cite{Shi:2023kid,Salmonson:1999es,AraujoFilho:2024mvz,Lambiase:2020iul}:
\begin{align}
x^{2}& = \sin^{2}\vartheta_{r}|_{\vartheta_{R}=0}\notag\\
&=1-\dfrac{\mathrm{R}^{2}}{r^{2}}\dfrac{\mathrm{g}_{tt}(r)}{\mathrm{g}_{tt}(\mathrm{R})}.
\end{align}

The evaluation of the total energy released in the vicinity of the compact source can be organized by first specifying the spatial domain influenced by its gravitational field and then accumulating the local deposition density throughout that region. The angular component plays a decisive role in this procedure, since its form is fixed by the geometry of the background. In practice, the metric governing the spacetime sets the structure of this contribution \cite{AraujoFilho:2025rzh,Shi:2023kid,Lambiase:2020iul,Shi:2025rfq}
\begin{align}
\dot{Q} & = \frac{\mathrm{d}\mathrm{E}}{\sqrt{-\mathrm{g}_{tt}(r)}\mathrm{d}t}\notag\\
&=\dfrac{84\zeta(5)\pi^{5}}{h^{6}}\mathrm{K}\, \mathrm{G}_{f}^{2} \, k^{9}
\left(\dfrac{7}{4}\pi a\,c\right)^{-\frac{9}{4}}
\Tilde{\mathcal{L}}_{\infty}^{\frac{9}{4}}\left[-\mathrm{g}_{tt}(\mathrm{R})\right]^{\frac{9}{4}}\notag\\
& \times 
\mathrm{R}^{-\frac{3}{2}}\int_{1}^{\infty}(x-1)^4\left(x^2+4x+5\right)\sqrt{\dfrac{\mathrm{g}_{rr}(y\mathrm{R})}{-\mathrm{g}_{tt}^9(y\mathrm{R})}}y^2\mathrm{d}y.
\end{align}

The quantity $\dot{Q}$ represents the amount of energy converted per unit time into electron--positron pairs through neutrino annihilation at a chosen radius \cite{Salmonson:1999es}. When this conversion rate grows sufficiently high, it can set off sizeable astrophysical activity driven by the newly created pairs. To isolate the role played by gravity in this mechanism, one may contrast the relativistic expression for $\dot{Q}$ with the formula obtained in a flat space treatment. This comparison makes it possible to identify how spacetime curvature alters the overall effectiveness of neutrino--powered energy release \cite{Salmonson:1999es,Shi:2023kid,Lambiase:2020iul,shi2022neutrino}

\begin{align}
\label{ratio_Q}
 \frac{\dot{Q}}{\dot{Q}_{\text{Newton}}} = 3\left[-\mathrm{g}_{tt}(\mathrm{R})\right]^{\frac{9}{4}}
& \int_{1}^{\infty}(x - 1)^{4}\left(x^{2} + 4x + 5\right) \\
& \times \sqrt{\dfrac{\mathrm{g}_{rr}(y\mathrm{R})}{-\mathrm{g}_{tt}^9(y\mathrm{R})}}y^2\mathrm{d}y.
\end{align}

where we have

\begin{align}
\mathrm{g}_{tt}(\mathrm{R})&= -\left(1 - \frac{2M}{\mathrm{R}}\right),\\
\mathrm{g}_{tt}(y\mathrm{R})&= -\left(1 - \frac{2M}{y\mathrm{R}}\right).
\end{align}

Furthermore, we also have the following relation
\begin{align}
x^{2}=1-\dfrac{1}{y^{2}}\frac{1-\dfrac{2M}{y\mathrm{R}}}{1-\dfrac{2M}{\mathrm{R}}}.
\end{align}

\begin{figure*}
\centering
\includegraphics[height=6.5cm]{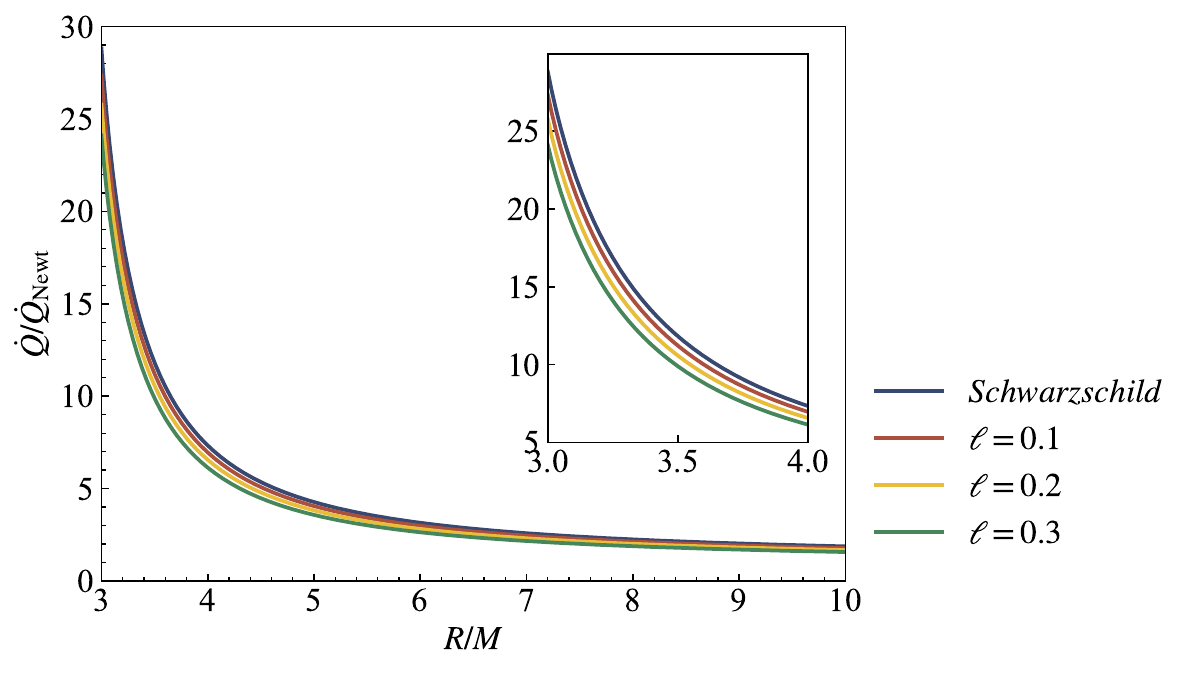}
\caption{$\dot{Q}/\dot{Q}_{\text{Newton}}$ plotted against $\mathrm{R}/M$ for several choices of the parameter $\ell$.}
\label{energgdfd}
\end{figure*}

\begin{figure*}
\centering
\includegraphics[height=6.5cm]{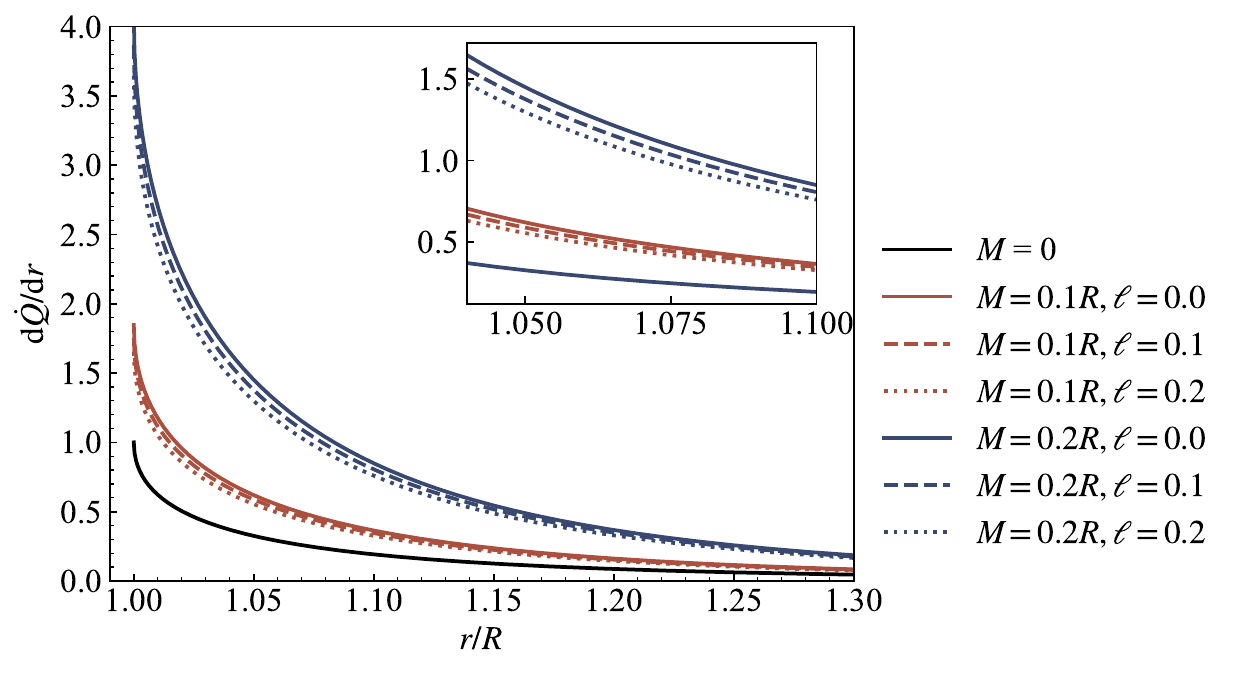}
\caption{Radial behavior of $\mathrm{d}\dot{Q}/\mathrm{d}r$ for several values of $M/\mathrm{R}$, including the Newtonian limit where $M=0$ yields $\mathrm{d}\dot{Q}/\mathrm{d}r=1$ at $r=\mathrm{R}$.}
\label{dQvfgdr}
\end{figure*}

The integrand appearing in Eq.~\eqref{ratio_Q} poses no numerical difficulties: it remains smooth throughout the domain $y\in[1,\infty)$ and exhibits no oscillatory behavior. The limiting values at the boundaries are straightforward to obtain, yielding $5\mathrm{R}^5\sqrt{1-\ell}/(\mathrm{R}-2)^5$ as $y\!\rightarrow\!1$ and vanishing as $y\!\rightarrow\!\infty$. Because of these properties, a standard Simpson–type routine provides a reliable evaluation of the integral, and the implementation available in \textit{scipy.integrate} was adopted for this purpose. The numerical tolerances were fixed at $1.49\times10^{-8}$ for both relative and absolute errors, leading to absolute uncertainties of order $10^{-11}$ (between $3.8\times10^{-11}$ and $4.8\times10^{-11}$) over the parameter range considered.

To examine how Lorentz--violating effects alter neutrino--antineutrino energy deposition, the calculation was reformulated using a Schwarzschild--like line element deformed by the Kalb--Ramond sector. This setup isolates the influence of the parameter $\ell$ on departures from general relativity. The resulting conversion–efficiency ratio from Eq.~(\ref{ratio_Q}) is displayed in Fig.~\ref{energgdfd} for several choices of $\ell$. All curves exhibit a reduction in the deposition rate when compared with the relativistic case, even as $\ell$ varies from $0$ to $0.3$, while their radial profiles remain qualitatively similar. Such behavior is consistent with the growing body of results on Kalb--Ramond--induced geometric modifications reported in recent studies \cite{Shi:2025rfq}.

%%%%%%%%%%%%%%%%%%%%%%%%%%%%%%%%%%%%%%%%%%%%%%%%%%%%%%%%%%%%%%%%%%%%%%%%%%%%%%%%%%%%%%%%%%%%%%%%%%%%%%%%%%%%%%%%%%%%%%%%%%%%%%%%%%%%%%%%%%%%%%%%%%%%%%%%%%%%%%%%%%%%%%%%%%%%%%%%%%%%%%%%%%%%%%%%%%%%%%%%%%%%%%%%%%%%%%%%%%%%%%%%%%%%%%%%%%%%%%%%%%%%%%%%%%%%%%%%%%%%%%%%%%%%%%%%%%%%%%%%%%%%%%%%%%%%%%%%%%%%%%%%%%%%%%%%%%%%%%%%%%%%%%%%%%%%%%%%%%%%%%%%%%%%%%%%%%%%%%%%%%%%%%%%%%%%%%%%%%%%%%%%%%%%%%%%%%%%%%%%%%%%%%%%%%%%%%%%%%%%%%%%%%%%%%%%%%%%%%%%%%%%%%%%%%%%%%%%%%%%%%%%%%%%%%%%%%%%%%%%%%%%%%%%%%%%%%%%%%%%%%%%

\section{Propagation phases and transition probabilities of neutrinos }

The motion of a neutrino associated with the $k$--th eigenstate in a spherically symmetric background can be formulated by starting from a variational description of particle dynamics in curved spacetime. Instead of introducing the metric first, the procedure begins with the Lagrangian approach developed in Ref.~\cite{neu18}, which is then applied to the geometry specified in Eq.~(\ref{maetriiccc}). This method produces the equations controlling the propagation of the neutrino mode through the spacetime by enforcing the corresponding action principle
\begin{align}
\mathcal{L}
& = -\frac{1}{2}  m_{k} \mathrm{g}_{tt}(r)\left(\frac{\mathrm{d}t}{\mathrm{d}\tau}\right)^2-\frac{1}{2}m_{k}\mathrm{g}_{rr}(r)\left(\frac{\mathrm{d}r}{\mathrm{d}\tau}\right)^2 \notag\\
& \quad -\dfrac{1}{2}m_{k}r^2\left(\frac{\mathrm{d}\theta}{\mathrm{d}\tau}\right)^2  
 -\frac{1}{2}m_{k}r^2\sin^2\theta\left(\frac{\mathrm{d}\varphi}{\mathrm{d}\tau}\right)^2.
\end{align}

When the motion is restricted to the equatorial region of the spacetime ($\theta=\pi/2$), only a few momentum components remain relevant. The analysis is organized by first defining the canonical momenta through
$p_{\mu}=\frac{\partial\mathcal{L}}{\partial(\mathrm{d}x^{\mu}/\mathrm{d}\tau)}$,
with $\mathcal{L}$ taken as the Lagrangian governing the trajectory and $\tau$ representing proper time. The mass of the neutrino mode under consideration is denoted by $m_{k}$. After imposing the symmetry constraint, the surviving components of $p_{\mu}$ follow directly, and their explicit expressions can be found in Refs.~\cite{neu60,Shi:2024flw}
\begin{align}
p^{(k)t} &= -m_{k}\mathrm{g}_{tt}(r)\frac{\mathrm{d}t}{\mathrm{d}\tau} = -E_{k}, \\
p^{(k)r} &= m_{k}\mathrm{g}_{rr}(r)\frac{\mathrm{d}r}{\mathrm{d}\tau}, \\
p^{(k)\varphi} &= m_{k}r^2\frac{\mathrm{d}\varphi}{\mathrm{d}\tau} = J_{k}.
\end{align}

Furthermore, it is important to highlight that the motion of a neutrino associated with the $k$--th mass eigenstate is constrained by the usual on--shell condition, as we should expect. Once the momentum components are specified, their contraction with the inverse metric must satisfy
$g^{\mu\nu}p_\mu p_\nu=-m_k^{\,2}$,
which enforces compatibility between the chosen trajectory and the spacetime background. This requirement ensures that the particle propagates in accordance with relativistic dynamics \cite{neu54}
\begin{align}
-m_{k}^2 =\mathrm{g}^{tt}p_t^2+\mathrm{g}^{rr}p_r^2+\mathrm{g}^{\varphi\varphi}p_{\varphi}^2.
\end{align}

In neutrino studies carried out far from strong gravitational fields, the description usually begins with flavor states rather than mass eigenstates, since weak interactions govern how these particles are produced and observed. Each flavor is a particular linear combination of mass modes, a fact emphasized in several earlier analyses \cite{neu63,neu62,Shi:2024flw,neu61}. Within such regions, where spacetime curvature plays no significant role, the propagation is commonly modeled through plane--wave treatments \cite{neu54,neu53}
\begin{align}
\ket{\nu_{\alpha}} = \sum \mathcal{U}_{\alpha i}^{*}\ket{\nu_{i}}.
\end{align}

A neutrino signal recorded at a detector originates from a flavor state—$\nu_e$, $\nu_\mu$, or $\nu_\tau$—since weak interactions select this basis. These flavor labels, indexed by $\alpha$, do not correspond to propagation eigenmodes; instead, each of them is a superposition of mass states connected through a unitary $3\times3$ matrix $\mathcal{U}$ \cite{neu41}. The dynamics are therefore more conveniently expressed in the mass representation, with states $\ket{\nu_i}$ following trajectories shaped by their individual masses. To describe the journey from the emission event at $\left(t_S,\bm{x}_S\right)$ to the detection point $\left(t_D,\bm{x}_D\right)$, one associates to each mass component a wave packet that propagates between these spacetime coordinates
\begin{align}
\ket{\nu_{i}\left(t_{D},\bm{x}_{D}\right)} = \exp({-\mathbbm{i}\Phi_{i}})\ket{\nu_{i}\left(t_{S},\bm{x}_{S}\right)}.
\end{align}
Consequently, the different mass components do not evolve identically during their journey from the emission point to the detector. Each one acquires its own propagation phase, which is obtained from the expression below:
\begin{align}
\Phi_{i}=\int_{\left(t_{S},\bm{x}_{S}\right)}^{\left(t_{D},\bm{x}_{D}\right)}\mathrm{g}_{\mu\nu}p^{(i)\mu}\mathrm{d}x^{\nu}.
\end{align}

In this description of neutrino propagation, one tracks how a state produced with flavor label $\alpha$ evolves as it travels to the detection event. Because the propagation is governed by the mass eigenstates rather than the flavor basis, the state reaching the detector may correspond to a different flavor $\beta$. The likelihood of observing this transition is quantified through the formula written below:
\begin{align}
\mathcal{P}_{\alpha\beta}
& = |\left\langle \nu_{\beta}|\nu_{\alpha}\left(t_{D}, \bm{x}_{D}\right)\right\rangle|^2 \\
& = \sum_{i,j} \mathcal{U}_{\beta i} \mathcal{U}_{\beta j}^{*} \mathcal{U}_{\alpha j} \mathcal{U}_{\alpha i}^{*}\,  \exp{[-\mathbbm{i}(\Phi_{i}-\Phi_{j})]}.
\end{align}

For neutrinos traveling in the geometry generated by a Kalb--Ramond black hole, the analysis is restricted to motion confined to the equatorial slice of the spacetime, $\theta=\pi/2$. With this symmetry imposed, the propagation of each mass mode acquires a phase determined by the following expression:
\begin{align}
\label{Pgefhi}
\Phi_{k} & = \int_{\left(t_{S},\bm{x}_{S}\right)}^{\left(t_{D}, \bm{x}_{D}\right)} \mathrm{g}_{\mu\nu} \, p^{(k)\mu}\mathrm{d}x^{\nu}\notag\\
& = \int_{\left(t_{S},\bm{x}_{S}\right)}^{\left(t_{D}, \bm{x}_{D}\right)}\left[E_{k}\mathrm{d}t - p^{(k)r}\mathrm{d}r-J_{k}\mathrm{d}\varphi\right] \notag\\
& = \pm\frac{m_{k}^2}{2E_0}\int_{r_{S}}^{r_{D}}\sqrt{-\mathrm{g}_{tt}\mathrm{g}_{rr}}\left(1-\dfrac{b^2|\mathrm{g}_{tt}|}{\mathrm{g}_{\varphi\varphi}}\right)^{-\frac{1}{2}}\mathrm{d}r.
\end{align}

In regimes where the spacetime curvature is negligible—specifically when $M/r \ll 1$—the quantity inside the integral of Eq.~(\ref{Pgefhi}) may be rewritten through a series expansion, as shown below
\begin{align}
&\quad\sqrt{-\mathrm{g}_{tt}\mathrm{g}_{rr}}\left(1-\dfrac{b^2|\mathrm{g}_{tt}|}{\mathrm{g}_{\varphi\varphi}}\right)^{-\frac{1}{2}}\notag\\
&\simeq\left(1-\dfrac{\ell}{2}\right)\left[\dfrac{r}{\sqrt{r^{2} - b^{2}}}-\dfrac{b^{2} M}{\left(r^{2} - b^{2}\right)^{\frac{3}{2}}}\right].
\end{align}
The expansion leads directly to a modified expression for the accumulated phase, which is written below:
\begin{align}
\Phi_{k} & = \dfrac{m_{k}^{2}}{2 E_{0}}\left(1-\dfrac{\ell}{2}\right)\Biggl[\sqrt{r_{D}^{2} - b^{2}}-\sqrt{r_{S}^{2} - b^{2}}\notag\\
&\quad+M\left(\dfrac{r_{D}}{\sqrt{r_{D}^{2} - b^{2}}}-\dfrac{r_{S}}{\sqrt{r_{S}^{2} -b^{2}}}\right)\Biggr].
\end{align}

In treating the propagation problem, the quantity that characterizes the neutrino’s typical energy at emission is expressed as
$E_0=\sqrt{E_k^{\,2}-m_k^{\,2}}$,
where the parameters $E_k$ and $m_k$ correspond to the energy and mass associated with the $k$--th eigenmode. The trajectory is labeled by the impact parameter $b$, following the convention adopted in Ref.~\cite{neu18}. As the particle moves through the curved geometry, its path reaches a turning point at the minimal radius $r_{0}$. When the gravitational field is weak, analytic approximations may be employed to determine this closest--approach radius
\begin{align}
\left(\dfrac{\mathrm{d}r}{\mathrm{d}\varphi}\right)_0=\pm\dfrac{\mathrm{g}_{\varphi\varphi}}{b^2}\sqrt{\dfrac{1}{-\mathrm{g}_{tt}\mathrm{g}_{rr}}-\dfrac{b^2}{\mathrm{g}_{rr}\mathrm{g}_{\varphi\varphi}}}=0.
\end{align}

The closest–approach radius $r_{0}$ follows from the conditions governing the orbit in the weak–field limit. In this approximation, the particle’s path imposes an equation whose solution yields the value of $r_{0}$
\ie
\label{r0}
r_{0} \simeq b - M.
\fe

The accumulated phase along the entire neutrino path—from emission, through the turning point, and onward to detection—can be obtained once the weak–field expression for $r_{0}$ in Eq.~(\ref{r0}) is incorporated into the integral of Eq.~(\ref{Pgefhi}). Combining these ingredients leads to the result displayed below:
\begin{align}
\label{pphhiii}
&\quad \Phi_{k}\left(r_{S}\to r_{0} \to r_{D}\right)\notag\\
&\simeq \frac{{m}_{k}^{2}}{2 E_{0}}\left(1-\dfrac{\ell}{2}\right)
\Biggl[\sqrt{r_{D}^{2}-r_{0}^{2}}+\sqrt{r_{S}^{2} -r_{0}^{2}}\notag\\
&\quad +M \left(\sqrt{\dfrac{r_{D} - r_{0}}{r_{D} + r_{0}}}+\sqrt{\dfrac{r_{S} - r_{0}}{r_{S} + r_{0}}}\right)\Biggr],
\end{align}
in which we have
\begin{align}
\Phi_{k}
& \simeq \frac{{m}_{k}^2}{2E_{0}}\left(1-\dfrac{\ell}{2}\right)
\Biggl[\sqrt{r_{D^{2}}-b^{2}}+\sqrt{r_{S^{2}}-b^{2}}\notag\\
& \quad + M\Biggl(\dfrac{b}{\sqrt{r_{D^{2}}-b^{2}}}+\dfrac{b}{\sqrt{r_{S^{2}}-b^{2}}}\notag\\
&\quad+\sqrt{\dfrac{r_{D}-b}{r_{D}+b}}+\sqrt{\dfrac{r_{S}-b}{r_{S}+b}}\Biggr)\Biggr].
\end{align}

To obtain a workable expression, the phase formula is next rewritten as a series in the small parameter ($b/r_{S,D}$). This procedure rests on the assumption that the impact parameter is negligible compared with the source and detector distances. Retaining contributions through ($\mathcal{O}(b^{2}/r_{S,D}^{2})$) leads to the approximation shown below:
\begin{align}
\label{Phi_k}
\Phi_k=\dfrac{m_{k}^{2}}{2 E_{0}}\left(1-\dfrac{\ell}{2}\right)(r_{D} + r_{S})\left(1-\dfrac{b^{2}}{2 r_{D} \, r_{S}}+\dfrac{2M}{r_{D} + r_{S}}\right).
\end{align}

An increase in the Lorentz--violating parameter $\ell$ systematically lowers the phase accumulated by propagating neutrinos. For the numerical illustration adopted here, the representative values $E_{0} = 10\,\mathrm{MeV}$, $r_{D}=10\,\mathrm{km}$, and $r_{S} = 10^{5}\, r_{D}$ were used. In the region surrounding the black hole, the spacetime curvature bends the trajectories, so neutrinos reaching the detector may follow several distinct paths. Determining the corresponding flavor--transition probabilities therefore requires evaluating the phase differences generated by these lensing--induced routes \cite{Shi:2024flw,AraujoFilho:2025rzh}

\begin{align}
\Delta \Phi_{ij}^{pq}
& = \Phi_i^{p} - \Phi_j^{q} = \Delta m_{ij}^2 A_{pq}+\Delta b_{pq}^2 B_{ij},
\end{align}

in which
\begin{align}
\label{Delta_m}
\Delta m_{ij}^2 & = m_{i}^{2} - m_{j}^{2},\\
\Delta b_{pq}^{2} & = b_{p}^{2} - b_{q}^{2},\\
A_{pq} & = \frac{r_{S} + r_{D}}{2 E_0}\left(1-\dfrac{\ell}{2}\right)\left[1+\dfrac{2M}{r_{D} + r_{S}}-\dfrac{\sum b_{pq}^{2}}{4 r_{D} r_{S}}\right],\\
B_{ij} & = -\frac{\sum m_{ij}^{2}}{8E_{0}}\left(1-\dfrac{\ell}{2}\right)\left(\frac{1}{r_{D}} + \frac{1}{r_{S}}\right),\\
\sum b_{pq}^2 & = b_{p}^{2} + b_{q}^{2},\\
\label{sum_m}
\sum m_{ij}^{2} & = m_{i}^{2} + m_{j}^{2}.
\end{align}

To keep track of the phases accumulated along the different neutrino routes, a superscript label is introduced so that $\Phi_{i}^{p}$ refers to the contribution of the $i$--th mass mode along the path indexed by $p$, each path being specified by an impact parameter $b_{p}$. The phase differences driving oscillation effects arise from the masses $m_{i}$, the mass–squared separations $\Delta m_{ij}^{2}$, and the geometric properties of the background. When the Lorentz--violating parameter is removed by taking $\ell \to 0$, the phase expression returns to the familiar form discussed in Ref.~\cite{neu53}.

The parameter $B_{ij}$ gathers all mass--dependent pieces, whereas the Lorentz--violating effects enter solely through the factor $A_{pq}$, which controls how the background geometry modifies both the size of the phase and the resulting oscillation pattern.

%%%%%%%%%%%%%%%%%%%%%%%%%%%%%%%%%%%%%%%%%%%%%%%%%%%%%%%%%%%%%%%%%%%%%%%%%%%%%%%%%%%%%%%%%%%%%%%%%%%%%%%%%%%%%%%%%%%%%%%%%%%%%%%%%%%%%%%%%%%%%%%%%%%%%%%%%%%%%%%%%%%%%%%%%%%%%%%%%%%%%%%%%%%%%%%%%%%%%%%%%%%%%%%%%%%%%%%%%%%%%%%%%%%%%%%%%%%%%%%%%%%%%%%%%%%%%%%%%%%%%%%%%%%%%%%%%%%%%%%%%%%%%%%%%%%%%%%%%%%%%%%%%%%%%%%%%%%%%%%%%%%%%%%%%%%%%%%%%%%%%%%%%%%%%%%%%%%%%%%%%%%%%%%%%%%%%%%%%%%%%%%%%%%%%%%%%%%%%%%%%%%%%%%%%%%%%%%%%%%%%%%%%%%%%%%%%%%%%%%%%%%%%%%%%%%%%%%%%%%%%%%%%%%%%%%%%%%%%%%%%%%%%%%%%%%%%%%%%%%%%%%%%%%%%%%%%%%%%%%%%%%%%%%%%%%%%%%%%%%%%%%%%%%%%%%%%%%%%%%%%%%%%%%%%%%%%%%%%%%%%%%%%%%%%%%%%%%%%%%%%%%%%%%%%%%%%%%%%%%%%%%%

\section{Neutrino path bending in curved spacetime }

\begin{figure}
    \centering
    \includegraphics[scale=0.45]{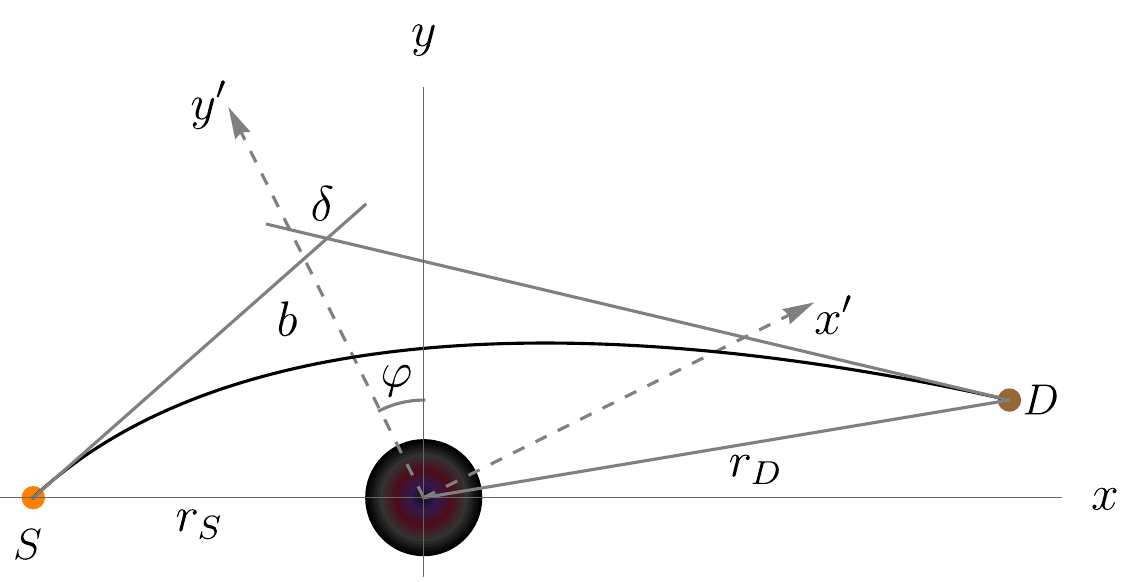}
    \caption{Neutrino trajectories distorted by mild gravitational lensing; $S$ denotes the emission point and $D$ the detection site.}
    \label{lensgigggg}
\end{figure}

A compact object with a strong gravitational field can deflect neutrinos so markedly that their motion departs from the radial direction, a consequence of gravitational lensing \cite{neu54}. Because of this bending, several distinct geodesics may link the same emission point to the detector $D$, as suggested in Fig.~\ref{lensgigggg}. When such a situation arises, the usual single–path treatment of flavor evolution is no longer adequate: the flavor state at $D$ must instead be written as a coherent sum over all admissible trajectories \cite{neu62,neu64,AraujoFilho:2025rzh,neu65,neu56,neu63,Shi:2024flw}:
\begin{align}
|\nu_{\alpha}(t_{D},x_{D})\rangle = \mathrm{N}\sum_{i}\mathcal{U}_{\alpha i}^{\ast}
\sum_{p} e^{- \mathbbm{i} \Phi_{i}^{p}}|\nu_{i}(t_{S}, x_{S})\rangle.
\end{align}

Each admissible neutrino path is indexed by $p$, and all of them terminate at the same detection point. Because the detector cannot distinguish which route was taken, the flavor observed at arrival results from the interference of the amplitudes carried along the different trajectories. The probability for a state produced as $\nu_{\alpha}$ to be identified as $\nu_{\beta}$ at the detector is therefore built from a coherent sum over these contributions, yielding the expression below \cite{neu65,Shi:2024flw,neu62,neu63,neu56,neu64}:
\begin{align}
\label{nasndkas}
\mathcal{P}_{\alpha\beta}^{lens} & = |\langle \nu_{\beta}|\nu_{\alpha}(t_{D}, x_{D})\rangle|^{2}\notag\\
& =|\mathrm{N}|^{2}\sum_{i, j}\mathcal{U}_{\beta i}\mathcal{U}_{\beta j}^{\ast}\mathcal{U}_{\alpha j}U_{\alpha j}^{\ast}\sum_{p, q}e^{\Delta \Phi_{ij}^{pq}}.
\end{align}
Within this framework, the factor ensuring proper normalization follows directly and is written below:
\begin{align}
|\mathrm{N}|^{2} = \left[\sum_{i}|\mathcal{U}_{\alpha i}|^{2}\sum_{p,q}\exp\left(-\mathbbm{i}\Delta \Phi_{ij}^{pq}\right)\right]^{-1}.
\end{align}

The quantity $\Delta\Phi_{ij}^{pq}$ introduced earlier governs how the various propagation routes interfere and thus determines the oscillation pattern when lensing is present. Its dependence on the masses $m_{i}$, the splittings $\Delta m_{ij}^{2}$, and the geometry encoded in Eq.~(\ref{nasndkas}) links the flavor transition mechanism to the curved background generated by the black hole. This structure resembles the formulations obtained for spherically symmetric metrics such as Schwarzschild spacetime \cite{AraujoFilho:2025rzh,neu53,Shi:2024flw}.

To assess how lensing affects the conversion between flavors, we now focus on the role played by the parameter $\ell$. For a two--flavor system, and assuming a weak--field regime, the transition probability from $\nu_{\alpha}$ to $\nu_{\beta}$ is evaluated by incorporating the spatial configuration of the source, the lensing region, and the detector \cite{neu53,neu65,Shi:2024flw,neu54}
\begin{align}
\label{as45dfdfd}
\mathcal{P}_{\alpha\beta}^{lens}
&=\left|\mathrm{N} \right|^2\biggl\{2\sum_i\left|\mathcal{U}_{\beta i}\right|^2\left|\mathcal{U}_{\alpha i}\right|^2\left[1+\cos\left(\Delta b_{12}^2B_{ii}\right)\right]\notag\\
&\quad+\sum_{i\neq j}U_{\beta i}\mathcal{U}_{\beta j}^*\mathcal{U}_{\alpha j}\mathcal{U}_{\alpha i}^*\notag\\
&\quad\times\left[\exp\left(-\mathbbm{i}\Delta m_{ij}^2 A_{22}\right) + \exp\left(-\mathbbm{i}\Delta m_{ij}^2 A_{11}\right)\right]\notag\\
&\quad+\sum_{i\neq j}\mathcal{U}_{\beta i}\mathcal{U}_{\beta j}^*\mathcal{U}_{\alpha j}\mathcal{U}_{\alpha i}^*\notag\\
&\quad\times \exp\left(-\mathbbm{i}\Delta m_{ij}^2A_{12}\right) \exp\left(-\mathbbm{i}\Delta b_{12}^2B_{ij}\right)\notag\\
&\quad+\sum_{i\neq j}\mathcal{U}_{\beta i}\mathcal{U}_{\beta j}^*\mathcal{U}_{\alpha j}\mathcal{U}_{\alpha i}^*\notag\\
&\quad\times\exp\left(\mathbbm{i}\Delta b_{21}^2B_{ij}\right)\exp\left(-\mathbbm{i}\Delta m_{ij}^2A_{21}\right)\biggr\}.
\end{align}

Equation~(\ref{as45dfdfd}) organizes the flavor--transition probability into several contributions, each identified by a particular combination of mass indices and trajectory labels. The term with $i=j$ isolates the evolution of a single mass mode, a situation in which no interference can arise. When $i\neq j$ but the propagation follows the same path ($p=q$), the expression instead takes into account the phase differences produced by distinct mass eigenstates sharing a common geodesic.

Additional interference patterns appear once both the mass labels and the path indices differ ($i\neq j$, $p\neq q$). These mixed contributions are separated into the sectors $p<q$ and $p>q$ to account for the asymmetry introduced by differing path lengths and the curvature of the background spacetime.

If only two--neutrino flavors are considered, the oscillation framework reduces to a simpler form. In that case, the link between flavor and mass bases is provided by a $2\times2$ unitary matrix characterized solely by the mixing angle $\alpha$ \cite{neu43}
\begin{align}
\label{U}
\mathcal{U}\equiv\left(\begin{matrix}
\cos\alpha&\sin\alpha\\
-\sin\alpha&\cos\alpha
\end{matrix}\right).
\end{align}

By inserting the matrix $U$ from Eq.~(\ref{U}) into the transition--probability formula of Eq.~(\ref{as45dfdfd}), one obtains the explicit expression governing the conversion of an electron neutrino into a muon neutrino, $\nu_{e}\!\rightarrow\!\nu_{\mu}$
\begin{align}
\label{proballl}
\mathcal{P}_{\alpha\beta}^{lens}
& =\left|\mathrm{N}\right|^2\sin^{2}2\alpha\notag\\
&\quad \times \biggl[\sin^2\left(\dfrac{1}{2}\Delta m_{12}^2A_{22}\right) + \sin^2\left(\dfrac{1}{2}\Delta m_{12}^2A_{11}\right)\notag\\
&\quad  +\dfrac{1}{2}\cos\left(\Delta b_{12}^2B_{22}\right) + \dfrac{1}{2}\cos\left(\Delta b_{12}^2B_{11}\right)\notag\\
&\quad - \cos\left(\Delta m_{12}^2A_{12}\right) \cos\left(\Delta b_{12}^2B_{12}\right)\biggr].
\end{align}

Once the matrix $U$ from Eq.~(\ref{U}) is specified and the path--dependent phase shifts are included, the factor required to normalize the total transition amplitude follows directly. Its explicit form is written below:
\begin{align}
\left|\mathrm{N}\right|^2& = \biggl[2\cos\left(\Delta b_{12}^2B_{11}\right)\cos^2\alpha + 2\notag\\
&\quad+ 2 \cos\left(\Delta b_{12}^2B_{22}\right) \sin^2\alpha\biggr]^{-1}.
\end{align}

%%%%%%%%%%%%%%%%%%%%%%%%%%%%%%%%%%%%%%%%%%%%%%%%%%%%%%%%%%%%%%%%%%%%%%%%%%%%%%%%%%%%%%%%%%%%%%%%%%%%%%%%%%%%%%%%%%%%%%%%%%%%%%%%%%%%%%%%%%%%%%%%%%%%%%%%%%%%%%%%%%%%%%%%%%%%%%%%%%%%%%%%%%%%%%%%%%%%%%%%%%%%%%%%%%%%%%%%%%%%%%%%%%%%%%%%%%%%%%%%%%%%%%%%%%%%%%%%%%%%%%%%%%%%%%%%%%%%%%%%%%%%%%%%%%%%%%%%%%%%%%%%%%%%%%%%%%%%%%%%%%%%%%%%%%%%%%%%%%%%%%%%%%%%%%%%%%%%%%%%%%%%%%%%%%%%%%%%%%%%%%%%%%%%%%%%%%%%%%%%%%%%%%%%%%%%%%%%%%%%%%%%%%%%%%%%%%%%%%%%%%%%%%%%%%%%%%%%%%%%%%%%%%%%%%%%%%%%%%%%%%%%%%%%%%%%%%%%%%%%%%%%%%%%%%%%%%%%%%%%%%%%%%%%%%%%%%%%%%%%%%%%%%%%%%%%%%%%%%%%%%%%%%%%%%%%%%%%%%%%%%%%%%%%%%%%%%%%%%%%%%%%%%%%%%%%%%%%%%%%%%%%

\section{Numerical investigation }

The evaluation of flavor oscillations in the black hole geometry requires the determination of the lensing probabilities introduced in Eq.~(\ref{proballl}). In the $(x,y)$ coordinate plane, the lens sits at the origin while the source and detector occupy the radial positions $r_{S}$ and $r_{D}$. To simplify the geometric description, the system is re--expressed in a rotated frame $(x',y')$, obtained by turning the original axes through an angle $\varphi$. This rotation yields the transformation rule given in \cite{neu53,Shi:2024flw}:
\begin{align*}
x' = x\cos\varphi + y\sin\varphi, \quad y' = -x\sin\varphi + y\cos\varphi .
\end{align*}

Choosing $\varphi=0$ fixes the geometry so that the source, lens, and detector lie along a single line in the plane. In this configuration, all three points occupy the same axis, which reduces the geometric complexity of the propagation setup and simplifies the treatment of the neutrino trajectories.

The bending experienced by a neutrino in the black hole spacetime—measured by the deflection angle $\delta$—depends directly on the impact parameter $b$, as shown in Refs.~\cite{neu53,Shi:2024flw}. In this arrangement, neutrinos originate at the source $S$, interact gravitationally with the Kalb--Ramond black hole acting as the lens, and eventually reach the detector $D$. The distances separating these points in the $(x,y)$ coordinate plane are labeled $r_{S}$ for the source--lens separation and $r_{D}$ for the lens--detector segment
\begin{align}
\label{delta}
\delta \sim\frac{y_{D}'- b}{x_{D}'}= \, \dfrac{4M}{b} -\dfrac{\pi\ell}{2}.
\end{align}

With the detector located at $(x_{D}',y_{D}')$ in the rotated coordinate frame and using the relation $\sin\varphi = b/r_{S}$, the deflection angle from Eq.~(\ref{delta}) can be rewritten in the following form:
\begin{align}
\label{asfffooooo}
&\quad\left( b y_{D} + 4M x_{D}-\dfrac{\pi\ell}{2}bx_{D}\right)\sqrt{1-\dfrac{b^{2}}{r_{S^{2}}}}\notag\\
& = b^2 \left(\dfrac{\pi\ell y_{D}}{2 r_{S}}+ \dfrac{x_{D}}{r_{S}}+1\right) - \dfrac{4Mby_{D}}{r_{S}}.
\end{align}

%%%%%%%%%%%%%%%%%%%%%%%%%%%%%%%%%%%%%%%%%%%%%%%%%%%%%%%%%%%%%%%%%%%%%%%%%%%%%%%%%%%%%%%%%%%%%%%%%%%%%%%%%%%%%%%%%%%%%%%%%%%%%%%%%%%%%%%%%%%%%%%%%%%%%%%%%%%%%%%%%%%%%%%%%%%%%%%%%%%%%%%%%%%%%%%%%%%%%%%%%%%%%%%%%%%%%%%%%%%%%%%%%%%%%%%%%%%%%%%%%%%%%%%%%%%%%%%%%%%%%%%%%%%%%%%%%%%%%%%%%%%%%%%%%%%%%%%%%%%%%%%%%%%%%%%%%%%%%%%%%%%%%%%%%%%%%%%%%%%%%%%%%%%%%%%%%%%%%%%%%%%%%%%%%%%%%%%%%%%%%%%%%%%%%%%%%%%%%%%%%%%%%%%%%%%%%%%%%%%%%%%%%%%%%%%%%%%%%%%%%%%%%%%%%%

\subsection{Two flavor case }

The study explored how the Kalb--Ramond black hole geometry alters neutrino oscillation behavior, with the Lorentz--violating parameter $\ell$ supplying the gravitational correction discussed earlier. Treating the detector as moving along a circular orbit around the Sun, its coordinates are written as $x_{D}=r_{D}\cos\varphi$ and $y_{D}=r_{D}\sin\varphi$. Within this setup, the quartic equation in Eq.~\eqref{asfffooooo} can be solved numerically in the equatorial plane, producing two positive real solutions, $b_{1}$ and $b_{2}$, for every angle $\varphi$. The quantities entering that polynomial are specified by Eqs.~\eqref{Delta_m}–\eqref{sum_m}, and the evaluation proceeds by inserting the chosen values of $m_{1}$, $m_{2}$, $b_{1}$, and $b_{2}$.

\begin{figure*}
\centering
\includegraphics[width=13.5cm]{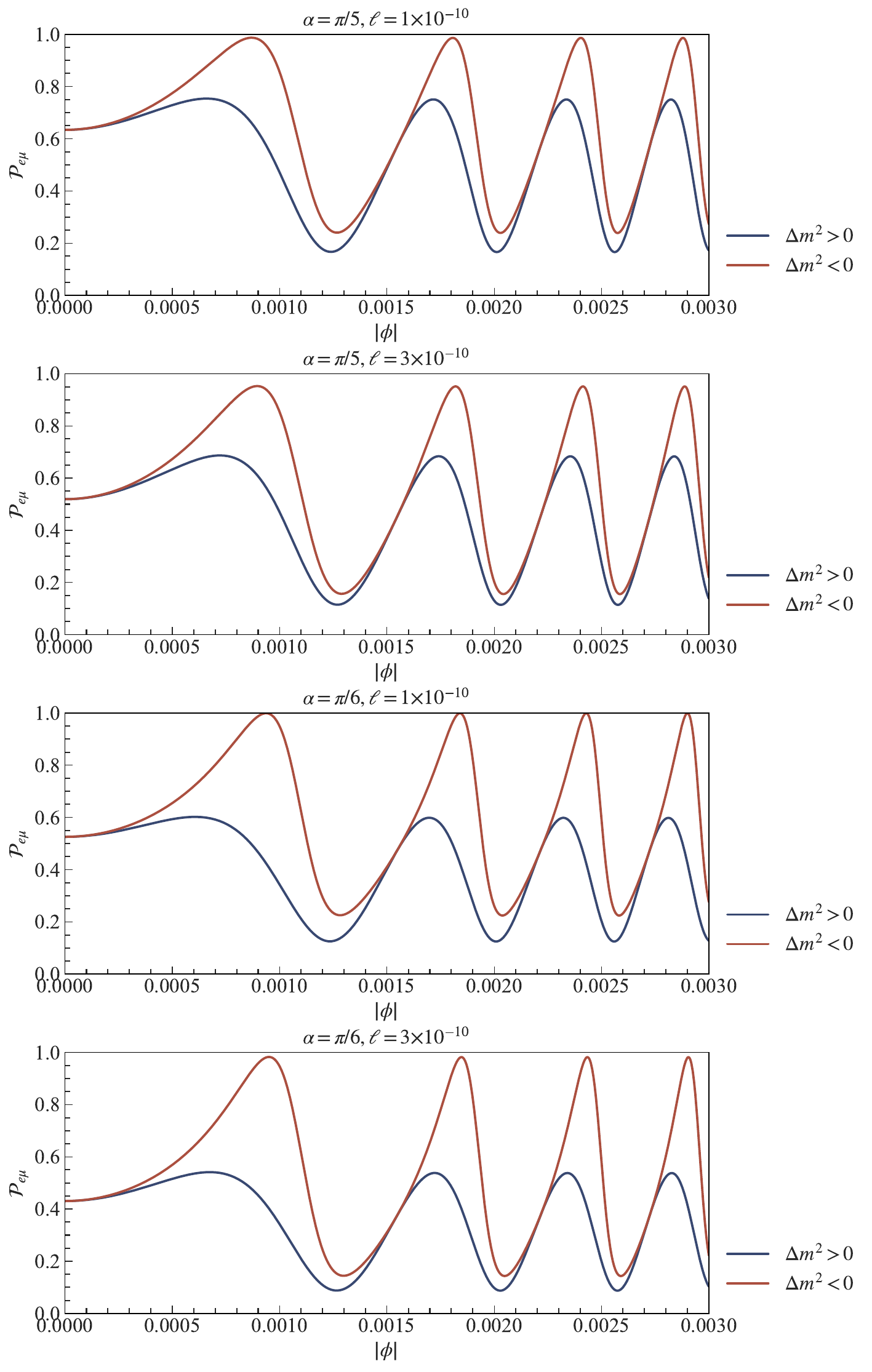}
\caption{\label{fig:prob1} Variation of $\nu_{e}\!\rightarrow\!\nu_{\mu}$ conversion with $\varphi$ for $\ell = 1\times10^{-10}$ and $3\times10^{-10}$ under different hierarchies and mixing angles.}
\end{figure*}

\begin{figure*}
\centering
\includegraphics[width=14.5cm]{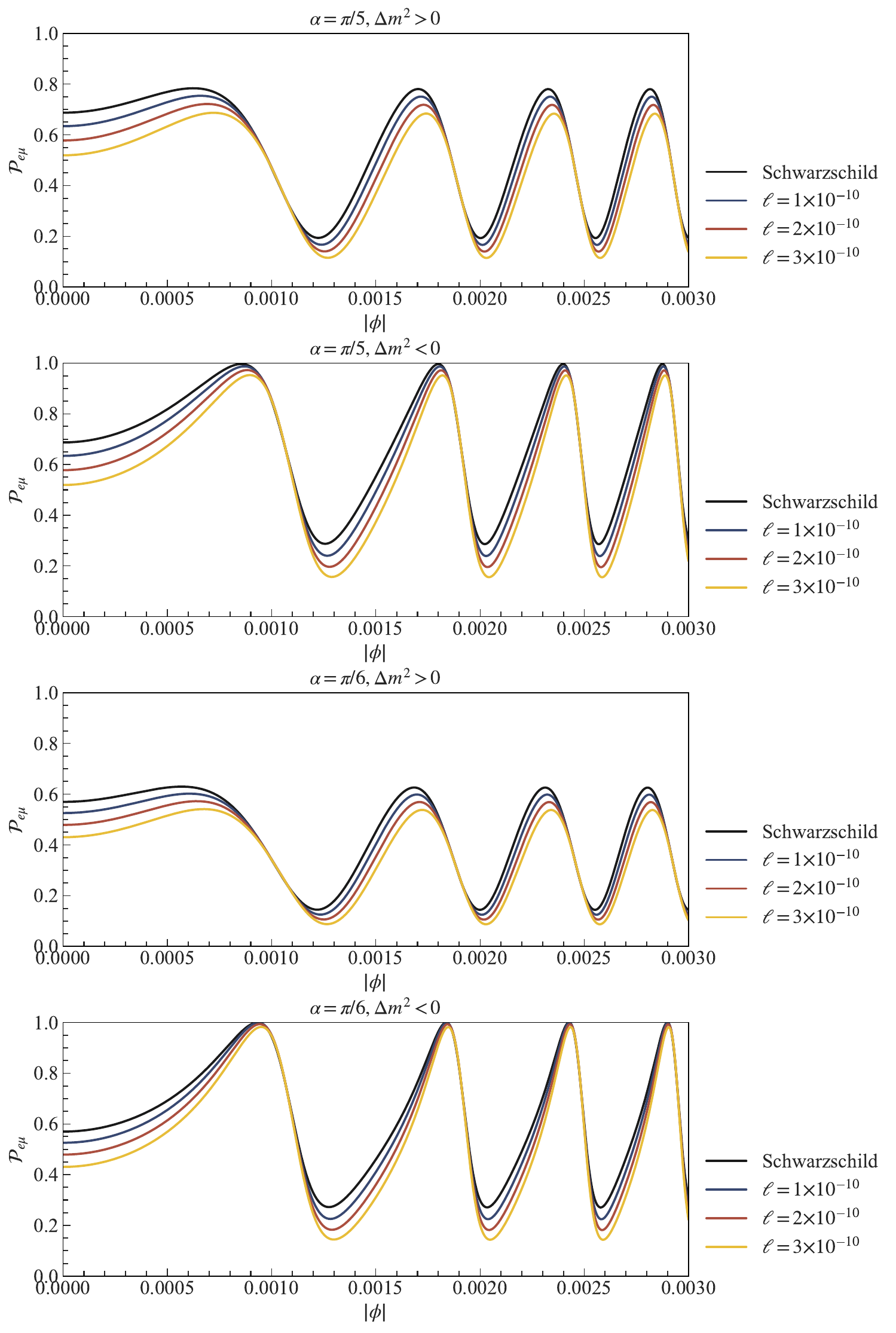}
\caption{\label{fig:prob2} Variation of $\nu_{e}\!\rightarrow\!\nu_{\mu}$ conversion with $\varphi$ for several $\ell$ values and mixing angles, evaluated for both mass hierarchies.}
\end{figure*}

\begin{figure*}
\centering
\includegraphics[width=16cm]{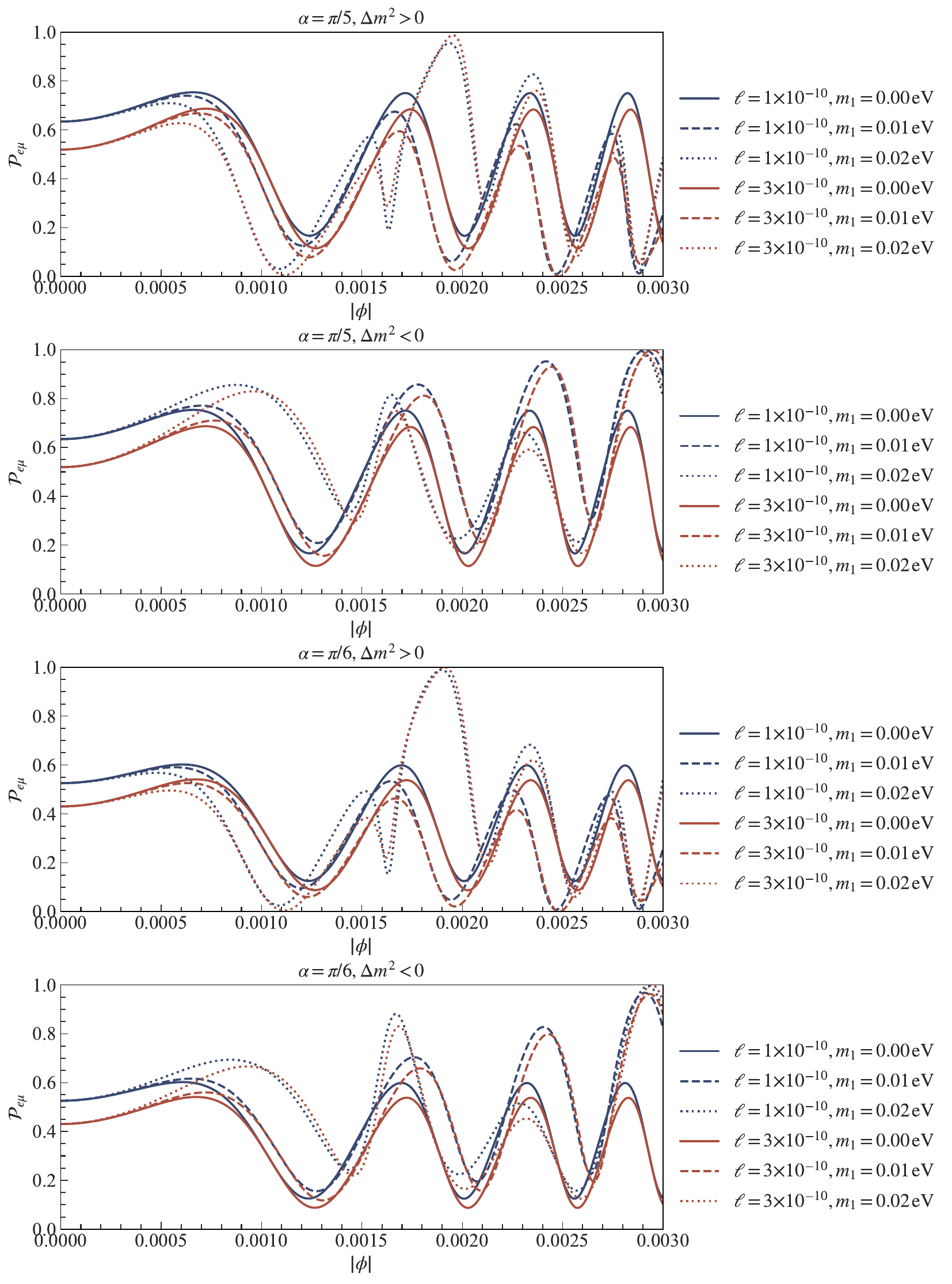}
\caption{\label{fig:prob3} Neutrino oscillation probability vs.\ $\varphi$ for normal ($\Delta m^{2}>0$) and inverted ($\Delta m^{2}<0$) hierarchies, shown for $\ell=1\times10^{-10}$ (blue) and $\ell=3\times10^{-10}$ (red).}
\end{figure*}

A comparison with the Schwarzschild limit ($\ell=0$) was then carried out. The oscillation patterns presented in Figs.~\ref{fig:prob1}, \ref{fig:prob2}, and \ref{fig:prob3} illustrate how variations of the azimuthal coordinate $\varphi$ and the mass–squared gap $\Delta m^{2}$ shape the resulting flavor--transition probabilities indeed.

Figure~\ref{fig:prob1} illustrates how the Kalb--Ramond background alters the behavior of the $\nu_{e}\!\rightarrow\!\nu_{\mu}$ transition, with the outcome depending strongly on the assumed mass ordering. For angles in the range $\varphi\in[0,0.003]$, the normal hierarchy ($\Delta m^{2}>0$) yields smaller conversion probabilities than the inverted hierarchy ($\Delta m^{2}<0$). Changing the mixing angle from $\alpha=\pi/5$ to $\alpha=\pi/6$ enhances the peak associated with the inverted scenario while suppressing the peak for the normal case, showing that the oscillation pattern becomes more sensitive to the parameter choices when the geometry departs from general relativity.

The effect of the Lorentz--violating parameter $\ell$ is displayed in Fig.~\ref{fig:prob2}. Increasing $\ell$ from $1\times10^{-10}$ to $3\times10^{-10}$ modifies the oscillation frequency only mildly, yet it reduces the overall magnitude of the transition probability. This behavior reflects a decrease in the accumulated phase, tied to the radial deformation of the spacetime.

Figure~\ref{fig:prob3} highlights the impact of the absolute mass scale. Raising the lightest mass $m_{1}$ from $0$ to $0.02,\mathrm{eV}$ suppresses the amplitude in the normal ordering while producing the opposite trend for the inverted ordering.

As established earlier in this work, the Lorentz--violating contribution arising in the Kalb--Ramond framework modifies only the radial part of the metric, $\mathrm{g}_{rr}$. This deformation elongates the neutrino trajectories and weakens the accumulated phase, producing shifts of roughly $20\%$ in the $\nu_{e}\!\rightarrow\!\nu_{\mu}$ conversion amplitude relative to the Schwarzschild prediction. The effect also depends on the mass ordering, generating variations that exceed $12\%$ across the azimuthal domain.

Situations in which neutrinos skirt the vicinity of a compact object provide natural settings where such differences become relevant. Examples include:
(i) emission from a core--collapse supernova whose proto--neutron star rapidly collapses to a black hole;
(ii) the MeV neutrino flux emerging from the hyper--accretion disk formed after a neutron--star merger; and
(iii) high--energy neutrinos produced along blazar jets and influenced by the supermassive black hole at the center.

Detectors such as Hyper--Kamiokande or DUNE would observe $\mathcal{O}(10^{5})$ MeV events for a source at 10 kpc, allowing statistical errors at the sub--percent level, while IceCube--Upgrade can probe the TeV--PeV regime. Since the predicted flavor modifications fall in the $10\!-\!20\%$ range, reaching an overall flavor--identification precision of $\lesssim5\%$ would enable a $3\sigma$ sensitivity—an accuracy expected from JUNO, DUNE, and Hyper--K for MeV neutrinos and from IceCube--Gen2 at high energies.

%%%%%%%%%%%%%%%%%%%%%%%%%%%%%%%%%%%%%%%%%%%%%%%%%%%%%%%%%%%%%%%%%%%%%%%%%%%%%%%%%%%%%%%%%%%%%%%%%%%%%%%%%%%%%%%%%%%%%%%%%%%%%%%%%%%%%%%%%%%%%%%%%%%%%%%%%%%%%%%%%%%%%%%%%%%%%%%%%%%%%%%%%%%%%%%%%%%%%%%%%%%%%%%%%%%%%%%%%%%%%%%%%%%%%%%%%%%%%%%%%%%%%%%%%%%%%%%%%%%%%%%%%%%%%%%%%%%%%%%%%%%%%%%%%%%%%%%%%%%%%%%%%%%%%%%%%%%%%%%%%%%%%%%%%%%%%%%%%%%%%%%%%%%%%%%%%%%%%%%%%%%%%%%%%%%%%%%%%%%%%%%%%%%%%%%%%%%%%%%%

\subsection{Three flavor case }

In order to assess the effect of Kalb--Ramond gravity on realistic neutrino transmission, we now expand our research to the complete three--flavor oscillation scenario. The oscillation parameters ($\theta_{ij}$, $\Delta m^2_{ij}$, and $\delta_{CP}$) were randomly selected within the $1\sigma$ confidence intervals of the NuFIT 5.0 global analysis in order to guarantee the physical robustness of our findings \cite{capozzi2014status,de2018status,esteban2019global}. 

\begin{figure*}
\centering
\includegraphics[width=17cm]{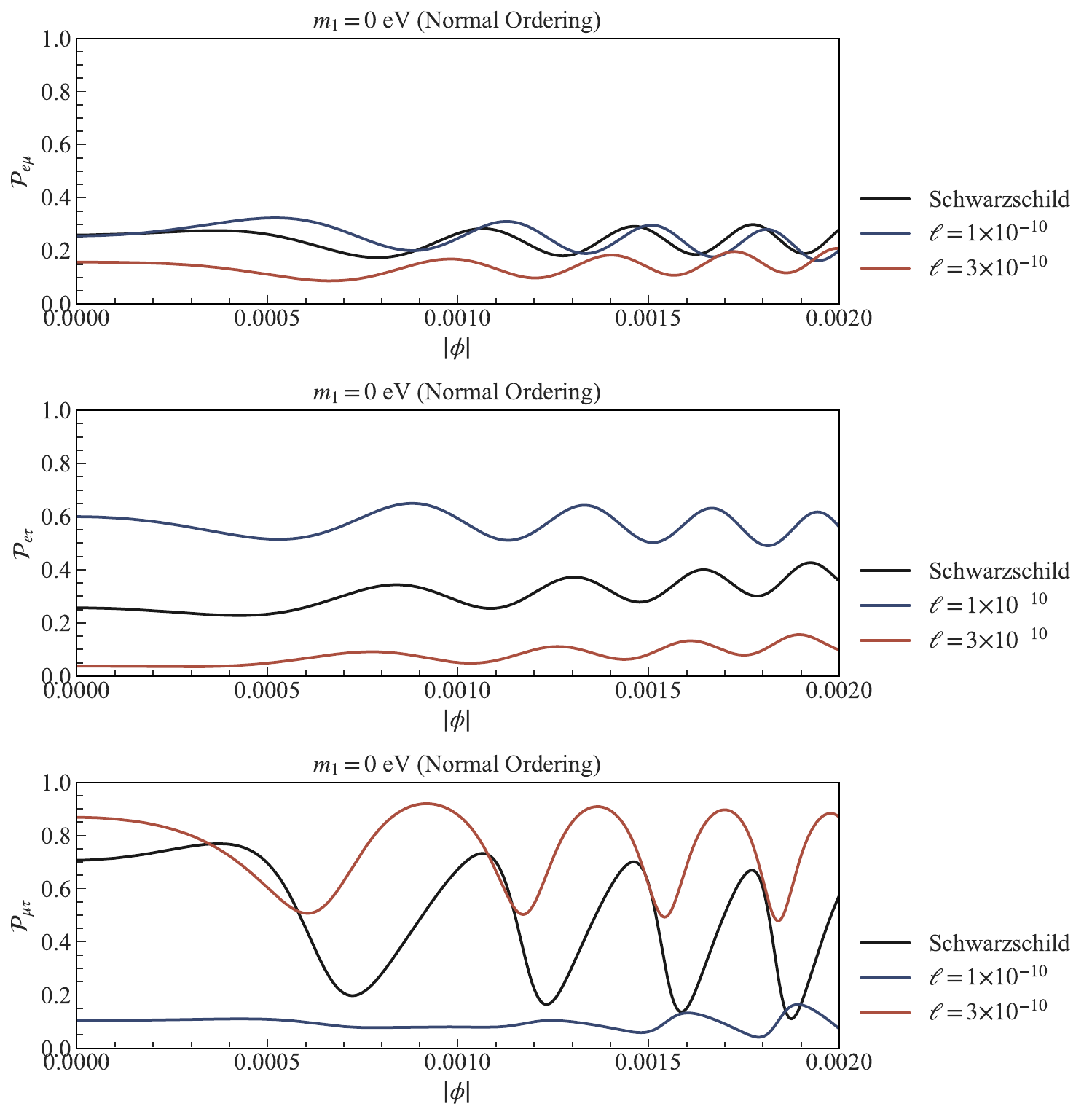}
\caption{\label{fig:prob4} The three oscillation probability of neutrinos is plotted as a function of the azimuthal angle $\varphi$ for normal hierarchies. The impact of Lorentz violation is illustrated through two parameter values: $\ell = 1\times10^{-10}$ (blue curves) and $\ell = 3\times10^{-10}$ (red curves). The oscillation parameters are sampled from NuFIT 5.0: $\Delta m_{21}^2=7.50\times10^{-5}\,\mathrm{eV^2}$, $\Delta m_{31}^2=2.51\times10^{-3}\,\mathrm{eV^2}$, $\theta_{12}=33.95^{\circ}$, $\theta_{13}=8.67^{\circ}$, $\theta_{23}=48.61^{\circ}$, and $\delta_{CP}=223.2^{\circ}$.}
\end{figure*}

Fig.~\ref{fig:prob4} displays the flavor conversion probabilities as a function of the azimuthal angle $\varphi$ for the case of Normal Ordering (NO). The specific set of oscillation parameters used for this simulation is: $\Delta m_{21}^2=7.5004\times10^{-5}\,\mathrm{eV^2}$, $\Delta m_{31}^2=2.5137\times10^{-3}\,\mathrm{eV^2}$, $\theta_{12}=33.9543^{\circ}$, $\theta_{13}=8.6672^{\circ}$, $\theta_{23}=48.6114^{\circ}$, and $\delta_{CP}=223.2054^{\circ}$ \cite{esteban2020fate,an2012observation}. The Lorentz--violating parameter $\ell$ causes a clear amplitude modulation, as seen in the Fig.~\ref{fig:prob4}. In contrast to a simple suppression, we find a distinctive rise in the oscillation envelope for the non--zero $\ell$ instances as the azimuthal angle $|\phi|$ grows, especially for $\ell=3\times 10^{-10}$ as shown by red curves. This suggests that an amplified interference pattern at greater deflection angles results from the accumulation of phase changes caused by the Kalb--Ramond geometry along the trajectory.

\begin{figure*}
\centering
\includegraphics[width=17cm]{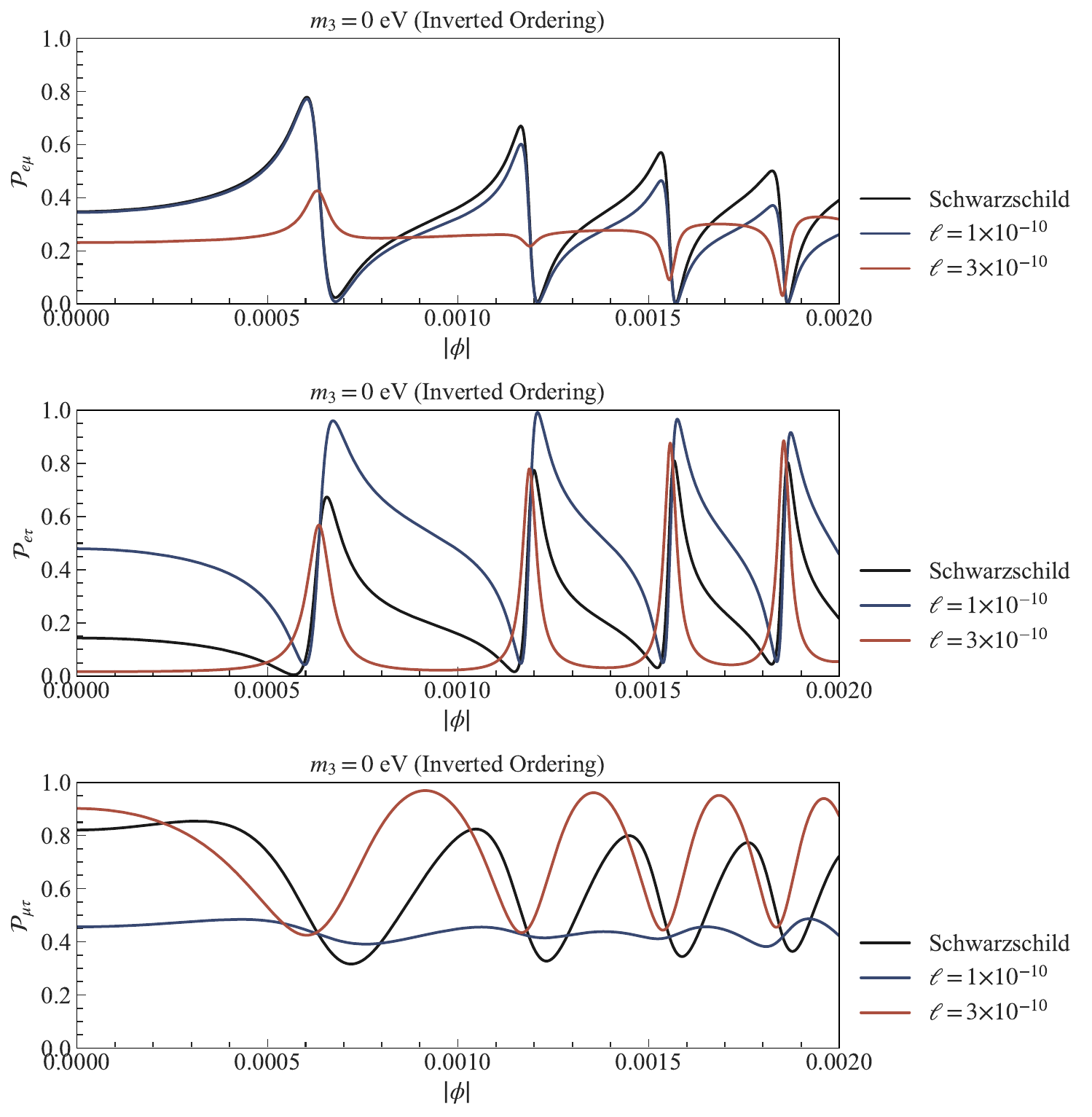}
\caption{\label{fig:prob5} The three oscillation probability of neutrinos is plotted as a function of the azimuthal angle $\varphi$ for inverted hierarchies. The impact of Lorentz violation is illustrated through two parameter values: $\ell = 1\times10^{-10}$ (blue curves) and $\ell = 3\times10^{-10}$ (red curves). The oscillation parameters are sampled from NuFIT 5.0: $\Delta m_{21}^2=7.53\times10^{-5}\,\mathrm{eV^2}$, $\Delta m_{32}^2=-2.52\times10^{-3}\,\mathrm{eV^2}$, $\theta_{12}=34.12^{\circ}$, $\theta_{13}=8.68^{\circ}$, $\theta_{23}=48.34^{\circ}$, and $\delta_{CP}=299.1^{\circ}$.}
\end{figure*}

In contrast, the Inverted Ordering (IO) scenario, presented in Fig.~\ref{fig:prob5}, exhibits distinct geometric resonances. The parameters selected for this case are: $\Delta m_{21}^2=7.5258\times10^{-5}\,\mathrm{eV^2}$, $\Delta m_{32}^2=-2.5198\times10^{-3}\,\mathrm{eV^2}$, $\theta_{12}=34.1187^{\circ}$, $\theta_{13}=8.6806^{\circ}$, $\theta_{23}=48.3405^{\circ}$, and $\delta_{CP}=299.1001^{\circ}$ \cite{esteban2020fate,an2012observation}. Sharp spikes appear in the probabilities $\mathcal{P}_{e\mu}$ and $\mathcal{P}_{e\tau}$ for minor Lorentz violations ($\ell=1\times 10^{-10}$, blue curves), suggesting constructive interference between the mass and geometry phases. In the $\mathcal{P}_{\mu\tau}$ channel, where the oscillation wavelength is significantly longer than in the Schwarzschild scenario, higher values of $\ell$ also result in a notable frequency redshift.

%%%%%%%%%%%%%%%%%%%%%%%%%%%%%%%%%%%%%%%%%%%%%%%%%%%%%%%%%%%%%%%%%%%%%%%%%%%%%%%%%%%%%%%%%%%%%%%%%%%%%%%%%%%%%%%%%%%%%%%%%%%%%%%%%%%%%%%%%%%%%%%%%%%%%%%%%%%%%%%%%%%%%%%%%%%%%%%%%%%%%%%%%%%%%%%%%%%%%%%%%%%%%%%%%%%%%%%%%%%%%%%%%%%%%%%%%%%%%%%%%%%%%%%%%%%%%%%%%%%%%%%%%%%%%%%%%%%%%%%%%%%%%%%%%%%%%%%%%%%%%%%%%%%%%%%%%%%%%%%%%%%%%%%%%%%%%%%%%%%%%%%%%%%%%%%%%%%%%%%%%%%%%%%%%%%%%%%%%%%%%%%%%%%%%%%%%%%%%%%%%%%%%%%%%%%%%%%%%%%%%%%%%%%%%%%%%%%%%%%%%%%%%%%%%%%%%%%%%%%%%%%%%%%%%%%%%%%%%%%%%%%%%%%%%%%%%%%%%%%%%%%%%%%%%%%%%%%%%%%%%%%%%%%%%%%%%%%%%%%%%%%%%%%%%%%%%%%%%%%%%%%%%%%%%%%%%%%%%%%%%%%%%%%%%%%%%%%%%%%%%%%%%%%%%%%%%%%%%%%%%%%%

\section{Conclusion}

Neutrino behavior in the Kalb--Ramond scenario proved to be strongly altered once spontaneous Lorentz violation, controlled by the parameter $\ell$, was introduced into the Schwarzschild--like geometry. The clearest signature appeared in the neutrino--antineutrino annihilation channel: after the energy--deposition formula was reconstructed on the modified metric, the total output for a source of radius $20\,\mathrm{km}$ and luminosity $10^{53}\,\mathrm{erg/s}$ increased by about five percent for $\ell=0.1$ and by roughly fourteen percent for $\ell=0.3$, even though the radial profile of $\mathrm{d}\dot Q/\mathrm{d}r$ preserved the same shape known from general relativity. This behavior originated from the deformation of $\mathrm{g}_{rr}$, which changed the radial gradient of the deposition rate.

The propagation sector was controlled by an oscillation phase multiplied by $(1-\ell/2)$,
$\Phi_{k}= \frac{m_k^2}{2E_0}\left(1-\frac{\ell}{2}\right)(r_D+r_S)!\left(1-\frac{b^2}{2r_D r_S}+\frac{2M}{r_D+r_S}\right)$, so the oscillation length increased whenever $\ell$ grew. Numerical studies showed that adjusting $\ell$ from $1\times10^{-10}$ to $3\times10^{-10}$ barely changed the azimuthal period but did reduce the $\nu_e\!\to\nu_\mu$ conversion amplitude by nearly twenty percent. This suppression resulted from the stretching of the optical metric rather than from any identifiable gravitational phase shift; for distant astrophysical emitters, the rapid phase accumulation was averaged out, leaving the amplitude as the physically relevant quantity.

The numerical analysis also indicated that the inverted mass ordering produced larger flavor--transition probabilities than the normal one when the lightest mass varied between $0$ and $0.02,\mathrm{eV}$. Phase calculations were carried out in Schwarzschild--like coordinates only to track radial dependence, since the chosen regime ($r_S = 10^5 r_D \gg 2M$) rendered near-horizon effects negligible. The phase definition itself stayed coordinate independent, as clarified in Ref.~\cite{neu54}, and the modifications originated from altered spatial distances and energies rather than from an additional “gravitational phase.”

The results pointed toward neutron--star mergers as plausible environments capable of exhibiting the oscillation distortions shown in Figs.~\ref{fig:prob1}–\ref{fig:prob3}, and the interval $\ell \ge 10^{-10}$ remained compatible with the sensitivity projected for IceCube-Gen2. Extensions of this work would require a quantum--field--theoretic treatment that includes spin--flip effects and decoherence due to curved backgrounds. Although decoherence was negligible in the weak--lensing regime, according to Ref.~\cite{neu53}, it created limitations for long--distance propagation when lensing generated mass--dependent interference.

Overall, the study revealed that spontaneous Lorentz violation in the Kalb--Ramond model increased the energy released by neutrino processes, expanded the oscillation scale, and produced flavor--transition modulations that depended on the mass hierarchy.

%%%%%%%%%%%%%%%%%%%%%%%%%%%%%%%%%%%%%%%%%%%%%%%%%%%%%%%%%%%%%%%%%%%%%%%%%%%%%%%%%%%%%%%%%%%%%%%%%%%%%%%%%%%%%%%%%%%%%%%%%%%%%%%%%%%%%%%%%%%%%%%%%%%%%%%%%%%%%%%%%%%%%%%%%%%%%%%%%%%%%%%%%%%%%%%%%%%%%%%%%%%%%%%%%%%%%%%%%%%%%%%%%%%%%%%%%%%%%%%%%%%%%%%%%%%%%%%%%%%%%%%%%%%%%%%%%%%%%%%%%%%%%%%%%%%%%%%%%%%%%%%%%%%%%%%%%%%%%%%%%%%%%%%%%%%%%%%%%%%%%%%%%%%%%%%%%%%%%%%%%%%%%%%%%%%%%%%%%%%%%%%%%%%%%%%%%%%%%%%%%%%%%%%%%%%%%%%%%%%%

\section*{Acknowledgments}
\hspace{0.5cm} A. A. Araújo Filho is supported by Conselho Nacional de Desenvolvimento Cient\'{\i}fico e Tecnol\'{o}gico (CNPq) and Fundação de Apoio à Pesquisa do Estado da Paraíba (FAPESQ), project numbers 150223/2025-0 and 1951/2025. V.B. Bezerra is partially supported by the Conselho
Nacional de Desenvolvimento Científico e Tecnológico (CNPq) grant number 307211/2020-
7. K.E.L.F  would like to thank the Paraíba State Research Support Foundation FAPESQ  for financial support.  A.R.Q work is supported by FAPESQ-PB. A.R.Q also acknowledges support by CNPq under process number 310533/2022-8.

\section*{Data Availability}

This manuscript has no associated data. (There is no observational data related to this article.)

\bibliographystyle{ieeetr}
\bibliography{main}

\end{document}